\documentclass[11pt]{article}
\usepackage{epsfig}
\usepackage{color}
\usepackage[max5]{authblk}

\def\be{\begin{eqnarray}}
\def\ee{\end{eqnarray}}
\def\bq{\begin{equation}}
\def\eq{\end{equation}}
\def\bse{\begin{eqnarray*}}
\def\ese{\end{eqnarray*}}

\newcommand{\bm}[1]{\mbox{\boldmath{$#1$}}}

\usepackage{fancyhdr}
\begin{document}
\title{{\bf A Millennium Bug Still Bites Public Health \\ -- An Illustration Using Cancer Mortality}
\author{Martina Fu $^1$,  David Todem $^2$, Wenjiang J. Fu $^2$, Shuangge Ma $^3$ \\
$^1$ Stanford University, $^2$ Michigan State University,  $^3$ Yale University }
\footnotetext{Corresponding authors: Wenjiang J. Fu, Department of Epidemiology and Biostatistics, Michigan State University, East Lansing, Michigan 48824 Tel: (517) 353 - 8623 ext 113. Email: fuw@msu.edu; and 
Shuangge Ma, Department of Biostatistics, School of Public Health, Yale University, New Haven, Connecticut 06520 Tel: (203) 785-3119. Email: shuangge.ma@yale.edu. }}

\date{}
\maketitle


\begin{center} {\bf \large  ABSTRACT}  \end{center}

Accurate estimation of cancer mortality rates and the comparison across cancer sites, populations
or time periods is crucial to public health, as identification of vulnerable groups who suffer the most from these diseases may lead to efficient cancer care and control with timely treatment. Because cancer mortality rate varies with age, comparisons require age--standardization using a reference population. The current method of using the Year 2000 Population Standard is standard practice, but serious concerns have been raised about its lack of justification. We have found that using the US Year 2000 Population Standard as reference overestimates prostate cancer mortality rates by 12--91\% during the period 1970--2009 across all six sampled U.S. states, and also underestimates case fatality rates by 9--78\% across six cancer sites, including female breast, cervix, prostate, lung, leukemia and colon-rectum. We develop a mean reference population method to minimize the bias using mathematical optimization theory and statistical modeling. The method corrects the bias to the largest extent in terms of squared loss and can be applied broadly to studies of many diseases. \\


\section*{Introduction}
Cancer is one of the leading causes of death in the United States and a major public health concern [1-5]. Cancer mortality rates are often reported in age-specific groups, making it difficult to extrapolate the overall mortality assessment or generate comparisons across populations \cite{Breslow}. Researchers often calculate a summary rate, such as the crude rate, which is an average of the age-specific mortality rates weighted to the proportions of age groups in the population. Such summary rates depend on the age-specific mortality rates and the population's age structure, where the latter may vary largely and cause unfair comparison among populations, presenting numerical illusion of large differences in the summary rate even as the age-specific mortality rates remain the same \cite{Doll1976}. This discrepancy motivated the direct age-standardization procedure for comparing mortality rates across populations [7-9] using a standard population as reference, such as the US Year 2000 Population Standard in current practice.

The age-standardization method calculates an age-adjusted rate using the age structure of a standard population to compare disease rates across populations or time periods. This method has been extensively adopted by the United States and world health agencies [4-5,10-12] following a memorandum from the Secretary of the U.S. Department of Health and Human Services in 1998 \cite{HHS1998} mandating the use of the US Year 2000 Population Standard to calculate age-adjusted mortality rate starting 1999 for more consistent reporting of mortality rate \cite{Anderson1998}. Accordingly, the US Year 2000 Population Standard or the Year 2000 World Standard Population has been used as reference in public health reports by many U.S. states [15-20] and worldwide agencies \cite{WHO}.

	Although age-standardization provides a way to compare disease rate among populations and has acknowledged merits [6-7,10-12], the current approach of using standard reference population also has caveats. It has been noted that choosing different reference populations may change the age-adjusted mortality rates dramatically and may also alter the ranking \cite{Doll1976,Robson2007}. As a result, the selection of standard population is still in debate \cite{Doll1976,WHO}. On one hand, selecting the Year 2000 Population Standard keeps the mortality rate adjustment consistent with a contemporary reference population \cite{HHS1998,Anderson1998}, making the comparison procedure easy to follow with uniformity \cite{Robson2007}. On the other hand, a study by the World Health Organization (WHO) pointed out ``There is clearly no conceptual justification for choosing one standard over another, hence the choice is arbitrary'' \cite{WHO}. Further, a health disparities study  attributed declining racial/ethnic and socioeconomic inequalities in health to the change of the reference population from Year 1940, 1970 to Year 2000, a ``statistical illusion'' due to the use of the Year 2000 Population Standard \cite{Krieger}. This illusive effect of the Y2K or millennium bug is not the result of a technical problem as in the computer programming case, but rather of a more difficult methodological one that requires theoretical research in quantitative science. The change of reference from US Year 1940 Population to US Year 2000 Population Standard may cause age-adjusted mortality rate to increase largely \cite{Anderson1998}, sometimes even more than doubling in size \cite{Sorlie1999}. The fact that the conclusion of mortality rate comparison depends on the choice of reference population creates confusions and misinterpretation \cite{Krieger}. Consequently, concerns about inadequate public health policymaking on healthcare, racial/ethnic and socioeconomic inequalities that result from improper comparisons need to be addressed urgently.

In this study, we investigated the issue of reference population selection using the Surveillance, Epidemiology and End Results (SEER) database \cite{SEER}. We analyzed prostate cancer mortality rates in six U.S. states (California, Massachusetts, Michigan, Missouri, New Jersey and New York) from 1970 to 2009 and also examined the U.S. case fatality rates in 2008 of six sites: female breast, cervix, prostate, lung, leukemia, and colon-rectum. We found that the age-standardization procedure using the US Year 2000 Population Standard as reference overestimated prostate cancer mortality rates in all six states and underestimated case fatality rates of all six cancer sites. This finding clearly indicates that bias has been introduced by the age-standardization procedure. To minimize the bias, we developed a mean reference population method to compare different populations. This method possesses a number of advantageous properties. First, the mean reference population minimizes the overall squared bias among all convex linear combinations. Second, by construction the mean reference population resembles the age profiles of all populations in comparison and thus represents them accurately while a standard population may present a largely different profile. Third, the existence and uniqueness of such a mean reference population is guaranteed by the mathematical optimization theory and can be computed using a mathematical quadratic programming method or a statistical sampling method. Fourth, the mean reference population method does not depend on the specifics of cancer mortality and can thus be applied to studies of incidence and mortality rates of other diseases. We show that the mean reference population method reduced to a large extent the overall bias associated with the use of the US Year 2000 Population Standard in the age-standardization procedure and yielded cancer mortality rate close to the crude rate.

\section*{Results}
By definition, the crude rate calculated with the total probability rule provides an unbiased estimate of the mortality rate \cite{Rosner}. To achieve fair comparison, however, a reference population is needed to remove artifacts due to varying age structure and bias is inevitably introduced by a reference population. See Material and Methods for more details and the definitions of various rates.

\begin{figure}
\centerline{\psfig{file=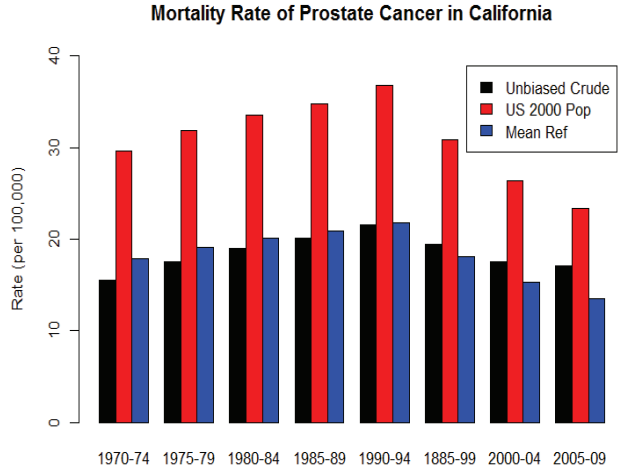,width=3.5in,height=3in}}
\caption{Comparison of age-adjusted mortality rates using the US Year 2000 Population Standard and the mean reference population to the crude rate of prostate cancer in the state of California during 1970-2009.}\label{fig1}
\end{figure}

\begin{figure}
\centerline{\psfig{file=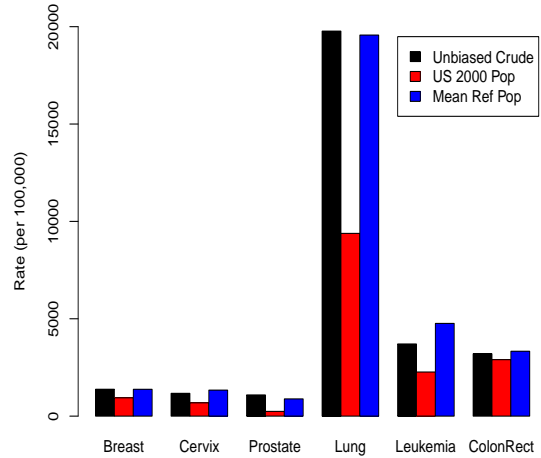,width=3in,height=3in}}
\caption{Comparison of age-adjusted case fatality rates using the reference of US Year 2000 Population Standard and the mean reference population to the unbiased crude rate of six cancer sites in 2008 in the United States.}\label{fig2}
\end{figure}

\begin{figure}
\centerline{\psfig{file=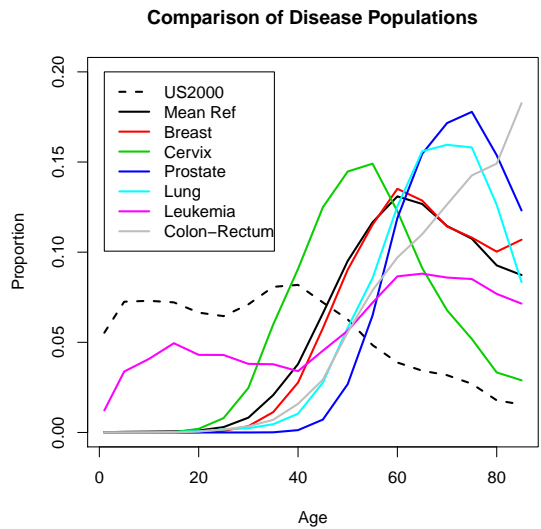,width=3in,height=3in}}
\caption{Comparison of age profile of the reference populations and the cancer patient populations of six sites. The profile of the mean reference population was similar to those of all cancer patient populations except for leukemia (which differed from others in early ages). In contrast, the decreasing trend of the US Year 2000 Population Standard differed largely from the increasing trend of the others, resulting in large bias in age-adjusted rate.}\label{fig3}
\end{figure}

Compared to the crude rate, the age-adjusted rate using the US Year 2000 Population Standard as reference overestimated prostate cancer mortality rate by 12\% to 91\% in all eight periods studied during 1970--2009, consistently in all six states (Tables 1). The cumulative rate, which takes the sum of the age-specific mortality rates from age 40 to 79 of each population \cite{Breslow}, yielded relative deviation of 1227\% to 3145\%, or 13 to 32 times that of the crude rate. In contrast, the mean reference population yielded age-adjusted rates much closer to the crude rates, with relative deviation between -25\% and 27\% and no systematic deviation towards either overestimation or underestimation. Fig. 1 illustrates the comparison in the State of California. Similar patterns were observed  in the other five states (Fig. S1 in Supplementary Material). Due to its large scale, the cumulative rate is not shown.

	Table 2 compares the observed number of deaths with the expected one in each population by the crude rate, the age-adjusted rates using the US Year 2000 Population Standard and using the mean reference population. The crude rate yielded exactly the same number of deaths as the observed, indicating unbiased estimation,  whereas the age-adjusted rate using the US Year 2000 Population Standard yielded much larger number of deaths than the observed, indicating biased estimation by  age-adjusted rate. Overall, the mean reference population yielded a more accurate estimation of the number of deaths than the US Year 2000 Population Standard. The expected number of deaths by the cumulative rate was not calculated because the cumulative rate is not a weighted average of the age-specific rates and thus is not a rate in a strict sense.

	The age-adjusted rate using the US Year 2000 Population Standard as reference underestimated the case fatality rate of all six cancer sites, as shown in Table 3. The relative deviations were -78\%, -52\%,   -42\%, -39\%,  -32\% and -9\% for cancer of prostate, lung, cervix, leukemia, female breast, and colon--rectum, respectively, indicating underestimation consistently across cancer sites. This discrepancy altered the ranking between leukemia and colon-rectum cancer. The cumulative rate overestimated the case fatality rate by 284--755\%. In contrast, the mean reference population yielded estimate much closer to the crude rate, with a mean deviation equal to 403.4, 90\% less than that by the US Year 2000 Population Standard (4305.1) and 99\% less than that by the cumulative rate (53351.3). Fig. 2 illustrates the comparison, again with the cumulative rate not shown because of its large scale.

	We then explored possible explanations for the large contrast of the age-adjusted rates between the two references of the US Year 2000 Population Standard and the mean reference population. We first examined the age profile of the reference populations using the case fatality study data and compared them to those of the six cancer patient populations. As shown in Fig. 3, five out of six cancers (except for leukemia) had virtually no patients before age 20, followed by a sharp increase between age 20, a peak between age 50 and 70 and a decline thereafter (except for colon-rectum cancer, which increased through age 85+). Mortality rates for leukemia were positive in early age, decreased slowly between age 15 and 40, and then increased thereafter. Overall, the mean reference population had an increasing trend similar to those of the six cancers, suggesting that it represented the cancer patient populations accurately. In contrast, the US Year 2000 Population Standard had an overall decreasing trend, staying large before age 20, peaking around 40 and then decreasing sharply thereafter. This sharp contrast suggests that the US Year 2000 Population Standard did not represent cancer patient populations, which explains why it yielded large deviation from the crude rate in comparing case fatality rates.

We then examined the weights of the reference populations. The mean reference population had a positive weight for each cancer site by construction and accurately represented the six cancer patient populations on the average.
$${\bm n}_{mean} = 0.5756{\bm n}_1 + 0.2172{\bm n}_2 + 0.0261{\bm n}_3 + 0.1757{\bm n}_4 + 0.0105{\bm n}_5 + 0.0048{\bm n}_6\, ,$$
where ${\bm n}_1, \ldots, {\bm n}_6$ are the population proportions of female breast, cervix, prostate, lung, leukemia, and colon-rectum, respectively. We further decomposed the US Year 2000 Population Standard by the six cancer patient populations using a linear regression model with no intercept for comparison with the above ${\bm n}_{mean}$.
$${\bm n}_{US2000} = - 1.1476{\bm n}_1 + 0.6034{\bm n}_2 + 0.3183{\bm n}_3 - 0.4891{\bm n}_4 + 1.6838{\bm n}_5 - 0.0017{\bm n}_6 \, .$$
The regression yielded three negative weights on breast, lung and colon-rectum cancers. In addition, the sum of the absolute values of all weights was 4.24, much greater than 1 as in the mean reference population. This result suggests that the US Year 2000 Population Standard was not ``close'' to a weighted average of the six cancer patient populations and thus was not a representative of them.

	We also examined the population profile of the six states during the  period 1970-2009 (Fig. S2-S7). Although the population in each state remained relatively stable, the effect of aging was observed by a shift of the peak from 1970 to 2009, indicating the change of population structure, the needs of age-standardization and the subsequent minimization of the overall bias as shown in Table~1 and Fig.~1 and S1.

\section*{Discussion}
	Accurate estimation of cancer mortality rate is a challenging task and has a major impact on cancer care and public health policymaking for cancer prevention, treatment and control [27-28]. The method of direct age-standardization has been studied for over a century \cite{Ogle1892}. Although concerns about the arbitrary selection of reference population are not new, they have become more urgent in recent years with the U.S. and world health agencies changing the reference population to reflect the contemporary, aging population structure. Such a change, though appealing in keeping mortality rate estimation consistent with a contemporary reference population, still lacks theoretical justification. As mentioned previously, the observed illusive effect on the declining racial/ethnic and socioeconomic inequalities raises more questions and demands renewed comparison in various aspects of disease incidence and mortality rates. The confusion caused by the selection of reference population is far from being clarified, and though a mandate of using a standard population as reference may help to streamline the age-standardization task, it may not help to adequately address the above concerns. Further theoretical research is urgently needed, more so than ever before.

	We have demonstrated that age-standardization using the US Year 2000 Population Standard overestimated prostate cancer mortality rate by up to 91\% and underestimated  case fatality rate by up to 78\%. Such large bias may result in confusion and misinterpretation of cancer mortality. For example, prostate cancer mortality may be misinterpreted as much higher than it actually was in all six states, and lung cancer case fatality rate may be misinterpreted as less than 10,000 per 100,000 person-year, while the actual rate was more than doubled ($>19,700$ per 100,000 person-year). Our observation is consistent with the concerns raised in the literature [7, 21, 22].

	Since the age-adjusted rate using the US Year 2000 Population Standard has been widely used in epidemiological studies and public heath reports, it has been regarded as the standard approach to comparing disease rates across populations. Many public health reports use it to generate disease rate estimation while acknowledging that the crude rate yields poor comparison with potential bias. For the first time, our study points out that the crude rate is unbiased and age-adjusted rates are biased. The use of a standard population in the direct age-standardization introduces bias and results in confusion and misinterpretation, as shown in Tables 1 and 2. The merit of the age-standardization is that it provides an equal footing for comparing disease rates among populations with different age structure, eliminating artifacts introduced by different population structure. Furthermore, we show in this paper that as long as one uses a common age structure to calculate age-adjusted rates, such equal footing is guaranteed. However, equal footing does not necessarily yield fair comparison because an age-structure may be in favor of one population over others. Hence, an equal footing age-structure may not be used as the only criterion for the fairness of comparison. This issue motivated us to search for a population that minimized the overall bias among all possible reference populations and led us to construct the mean reference population based on the populations in comparison.  Our mean reference population method not only provided a common population structure for comparison but also minimized the overall bias.

	Although the effect of age-standardization differed in the two cancer mortality studies,  overestimation of mortality rate and underestimation of case fatality rate, both showed a consistent lack of calibration by age-adjusted rate, indicating the need for improvement. Our mean reference population method minimizes the overall bias, and may also help to address the issue of arbitrary selection of reference population raised in the WHO report \cite{WHO}.

	Although cancer case fatality rates may be inaccurate due to lead bias in cancer diagnosis, the principle of our analysis remains the same, and the underestimation of case fatality rates by the US Year 2000 Population Standard remains a valid conclusion. For example, take prostate cancer, a disease with a late onset at age 35 or older (Table S1). A major proportion (45\%) of the US Year 2000 Population Standard is younger than age 35, a population in which prostate cancer rarely develops (assuming equal distribution by age between males and females, thus gender effect need not be considered). Hence, the age-standardization by the US Year 2000 Population Standard only accounts for 55\% weight on death from prostate cancer, largely underestimating the case fatality rate. Similar explanation holds for other cancers.

	We appreciate that although the mean reference population provides a unique reference population and minimizes the overall bias of age-adjusted rate among given populations, it does not remain the same  in different studies and has to be constructed for each comparison study. This leads to technical inconvenience in comparing disease rates. To resolve this issue, we plan to provide a computer software package to  implement the procedure, for which computation of optimal weights and storage of  population proportions of racial and sex groups in geographic locations is inexpensive.

	We conclude that direct age-standardization using a standard population may lead to inaccurate estimation and incorrect interpretation, resulting in confusions and inappropriate decision and policy making. It is hoped that the mean reference population method may lead to improved cancer patient care and efficient healthcare management. Furthermore, since the method relies on no specification of cancer mortality rates, it applies broadly to studies comparing incidence or mortality rates of a wide range of diseases in varied countries and geographic regions.

\section*{Materials and Methods}
\subsection*{Data}
	Prostate cancer mortality rates and population proportions of five year age groups and five year periods during the years 1970 - 2009 were generated for six US states (California, Massachusetts, Michigan, Missouri, New Jersey, and New York), from the Surveillance, Epidemiology and End Results (SEER) database \cite{SEER} using the SEER*Stat software version 8.0.4 \cite{SeerStat}. These six states were selected because they used the age--standardization method to generate  state public health reports of cancer mortality [15-20]. The SEER database consists of cancer incidence and mortality data of U.S. cancer registries in a growing number (nine and more) of metropolitan areas since the 1970s. Hence research results based on the SEER database are often interpreted as the results for the United States.

	The U.S. case fatality rates in 2008 were generated using the SEER database and the Cancer Prevalence database of the NCI/NIH \cite{CanQues}. We first generated the U.S. cancer mortality rates for each cancer site with the SEER*Stat software, and then calculated the case fatality rates using the prevalence of each cancer site estimated by the software CanQues Version 4.2 \cite{CanQues}. See Supplementary Material for details.

\subsection*{Methods}
{\bf Age-standardization for  comparing mortality rate across populations  \hspace{.2em}}
Cancer mortality rate varies with age (Tables S3, S5, S7, S9, S11 and S13), and a summary rate (e.g. the crude rate) is often preferred to a sequence of age-specific rates in comparing the mortality [6-7,10-12]. The age-standardization yields an age-adjusted rate,  a weighted average of the age-specific rates $m_a$ with selected weights $n_a$,
\begin{equation}
r_{adj}=\sum_{a=1}^A m_an_a \, .
\end{equation}
{\bf Age-adjusted rates \hspace{.2em}} Four  rates were calculated to summarize the age-specific mortality rates for comparison, the crude rate, the cumulative rate,  age-adjusted rates using the US Year 2000 Population Standard and using the mean reference population. They were calculated as follows for given age-specific mortality rate $m_{ia}$, population proportion $n_{ia}$ of population $i$, and population proportion of a standard population $n_{0a}$.
	The crude rate $r_{crude\; i}$  of prostate cancer mortality in population $i$ in each state and each period during 1970--2009 was calculated using the total probability rule \cite{Rosner}
\[r_{crude\; i} = \sum_{a=1}^A m_{ia}n_{ia}, \]
where the weights were its own population proportions $n_{ia}$. The US crude case fatality rate of each cancer site in 2008 was calculated similarly. The cumulative rate $r_{cumul\; i}$  of each population $i$ was calculated to be the sum of the age--specific mortality rates from age 40-44 to age 75-79 for each state and each period \cite{Breslow}, $r_{cumul\; i} = \sum_{a=40}^{75}m_{ia}$, where the weights $n_{ia}=1$ for $a=$ 40, 45, $\ldots,$ 75, and  0 otherwise. Each age-adjusted rate was calculated following the direct age-standardization procedure in equation (1) see \cite{Breslow}. The US 2000 age-adjusted rate $r_{US2000}$ was calculated  using the US Year 2000 Population Standard proportions $n_{US2000\; a}$  as weights, and the  mean reference rate $r_{mean}$  was calculated using a mean reference population proportion $ n_{mean\; a}$ as weights, where the mean reference population was constructed using a convex linear combination of the proportions $n_{ia}$ of the populations in comparison. See Statistical Modeling below for details.

\noindent
{\bf Unbiased estimation of overall mortality by crude rate \hspace{.2em}}
Assume we have $A$ age groups.  The $A$ age-specific mortality rates of each population form an $A$-vector ${\bm m}=(m_1, \ldots, m_A)^T$. Also assume the proportion of a given population with the disease is ${\bm n}=(n_1,\ldots, n_A)^T$ for $A$ age groups,  $\sum_{a=1}^A n_a=1$. Assume that a study has $p$ populations, and each population has a mortality rate vector ${\bm m}_i$ and a corresponding population proportion vector  ${\bm n}_i$, $i=1, \ldots, p$.  The mortality rate of each population in a given period of time is defined in \cite{Breslow} as
{
\[r = \frac{\rm Total \; number \;of\; deaths \; during \;given \; period }{\rm Mid \; point \;  population \;  \times \; duration \; of \; period  }.\]
}
By the total probability rule \cite{Rosner}, the crude mortality rate of each population is estimated with $ \hat r = \sum_{a=1}^A m_an_a = {\bm m}^T{\bm n} $, an inner product of the two vectors ${\bm m}$ and ${\bm n}$, i.e.  the crude mortality rate is a weighted average of the age-specific rates. It is shown below that the crude mortality rate provides an unbiased estimate of the overall mortality rate over one year period for each population. This explains why the crude rate yielded the same number of deaths as the observed (Table 2).
{\baselineskip = 0.8\baselineskip
\begin{eqnarray}
\hat r &=&\sum_{a=1}^A m_an_a
  =\sum_{a=1}^A \frac{{\rm Number \;of\; deaths \;in \;} a{\rm -th \; age \; group}}{{\rm Population\; in \;} a {\rm -th \; age\; group} \times 1 \; {\rm year } }
 \times \frac{{\rm Population\; in\;} a {\rm -th \; age\; group}}{\rm Total\; population}\nonumber \\
&=& \sum_{a=1}^A\frac{{\rm Number \; of \; deaths \; in} \; a {\rm -th \; age\; group}}{{\rm Total \; population}\; \times 1 \; {\rm year}}
=\frac{\rm Total \; number \; of \; deaths}{{\rm Total \; population}\; \times 1 \; {\rm year}}.
\end{eqnarray} }
Thus the unbiased estimates of the mortality rates of the $p$ populations are
\begin{equation}
\hat r_1={\bm m}_1^T{\bm n}_1 \; , \hspace{1em} \cdots,  \hspace{1em}   \hat r_p={\bm m}_p^T{\bm n}_p \; .
\end{equation}
{\bf Comparing multiple populations with age-adjusted mortality rates\hspace{.5em}}
Equation (3) provides an unbiased estimate of the overall mortality rate for each population. However, the rates so generated cannot provide a fair comparison across populations as different populations may have different age structures. With a late-onset disease, it is very likely that an older population yields a higher mortality rate than a younger population, even if the two populations have the same age-specific mortality rates \cite{Doll1976,NCHS1998}. For fairer comparison, a direct age-standardization procedure was studied \cite{Doll1976}, in which an age-adjusted rate was calculated based on age-specific rates ${\bm m}$ of a population in comparison and age structure ${\bm n}$ of a standard population, such as the US Year 2000 Population Standard [4,5,10] or the WHO World Standard Population \cite{WHO}. The age-adjusted mortality rate was calculated as $R_i={\bm m}_i^T{\bm n}$ for the $i$-th population. The expected number of deaths was thus calculated by multiplying the rate $R_i$ by the total population $N_i$ of population $i$, $R_iN_i$. The calculation of expected number of deaths allowed comparison of the rate among populations, and more importantly allowed comparison of the expected number of deaths of each population to the observed, assessing the bias of each rate.

\noindent
{\bf Bias introduced by age-standardization \hspace{.2em}}
We made  four observations below. \\
1) The crude rate in equation (3) provides unbiased estimation for each population. 2) An age-adjusted rate using a reference population ${\bm n}$ may deviate from the crude rate and introduce bias. The deviation is calculated with $ b_i = {\bm m}_i^T {\bm n} - {\bm m}_i^T {\bm n}_i = {\bm m}_i^T ({\bm n}-{\bm n}_i) $. 3) The bias is often inevitable in comparison among multiple populations. $ b_i\neq 0$ unless the difference vector ${\bm n}-{\bm n}_i$ between the reference population and the $i$-th population is perpendicular to the rate vector ${\bm m}_i$. Since multiple populations are often compared in a given study, the chance that one single reference population ${\bm n}$ makes the deviation of all populations equal to $0$ is extremely small because ${\bm n} = (n_1,\ldots,n_A)^T$ needs to satisfy conditions $n_a\geq 0$ and $\sum_{a=1}^A n_a=1$. 4) It is thus desirable to find a reference population ${\bm n}$ to minimize the overall bias for all $p$ populations \cite{Doll1976}.

	The difference between age-adjusted and crude rates represents an estimate of bias caused by using a reference population. We define a relative deviation to be the deviation of an age-adjusted rate ($R_{\rm adjust}$) as a percentage of the crude rate ($R_{\rm crude}$),
 \[{\rm Reletive\; Deviation} = \frac{R_{\rm adjust} - R_{\rm crude}} {R_{\rm crude}}\times 100\% .\]
A positive value indicates overestimation and a negative one indicates underestimation.

To calibrate the age-adjusted rate with a  mean reference population, we took a weighted average of the proportions of all populations in comparison, i.e. the proportions of six US cancer patient populations (Table S1) or the population proportions in eight periods of each state (Tables S4, S6, S8, S10, S12 and S14). We selected the weights by minimizing the total deviation, which is defined as the sum of squares of the deviations across all populations in comparison. By definition, the mean reference population is optimal in calibrating the age-adjusted mortality rate.

	We compared the age-adjusted rate with the crude rate, using the US Year 2000 Population Standard or the mean reference population as reference. We used the mean deviation (averaged over all populations in comparison) to assess each age-adjusted rate. See equation (8) below for the definition of mean deviation. We also compared age profile of the US Year 2000 Population Standard and the mean reference population with the populations in comparison, as shown in Fig. 3.

\noindent
{\bf Criteria in searching for a reference population\hspace{.2em}}
We searched for a reference population ${\bm n}$ by minimizing the total squared deviation of $p$ populations
\begin{equation}
\min_{\bm n} \left\{\sum_{i=1}^p b_i^2 \equiv \sum_{i=1}^p [{\bm m}_i^T ({\bm n}-{\bm n}_i)]^2 \right\}.
\end{equation}
Technically, a reference population ${\bm n}$ needs to satisfy $n_a\geq 0$ and $\sum_{a=1}^A n_a=1$ for population proportion. However, these conditions may not be enough to ensure that the reference population is ``close'' to or representative of the given populations ${\bm n}_1, \ldots, {\bm n}_p$. In order to make the reference population represent the given populations, we further required that the search for the reference population be conducted among weighted averages of the given populations. Mathematically they are convex linear combinations of these populations ${\bm n} = t_1{\bm n}_1 + \ldots + t_p{\bm n}_p$ with $t_i\geq 0$ for $i=1, \ldots, p$ and $\sum_{i=1}^p t_i=1$. Such a convex linear combination ensured that the target reference population ${\bm n}$ was representative and retained the characteristics of the $p$ populations. Therefore, our objective was to search for a set of weights $t_i\geq 0$ for $i=1, \ldots, p$ satisfying  $\sum_{i=1}^p t_i=1$ such that the linear combination ${\bm n}$ minimized the total deviation in equation (4). This approach formulated the objective into a mathematical optimization problem \cite{Lee2005,JensenIneq}.

\noindent
{\bf Optimization by quadratic programming\hspace{.2em}}
Let $f({\bm t}) = \sum_{i=1}^p  b_i^2$ be the total deviation to be minimized. Then
\begin{equation}
 f({\bm t}) = \sum_{i=1}^p ({\bm m}_i^T {\bm n}- r_i)^2 = \|M^T{\bm n}-{\bm R}\|^2\,,
\end{equation}
where $M=({\bm m}_1, \ldots, {\bm m}_p)$ is an $A\times p$ matrix. The column vectors ${\bm R} = (r_1, r_2, \ldots, r_p)^T$ and  ${\bm t} = (t_1,\ldots, t_{p})^T $. $\|\cdot\|$ is the Euclidean norm. Since ${\bm n} = t_1{\bm n}_1 + \ldots + t_p {\bm n}_p $, $f({\bm t})$ is a quadratic function of $t_1, \ldots, t_p$, and can be minimized in a compact domain. Since the constraints $t_i\geq 0$ for $i = 1, \ldots, p$ and $\sum_{i=1}^p t_i=1$ form a simplex ${\cal D}$ in a $p$-dimensional space, which is compact, the function $f({\bm t})$ can be minimized in ${\cal D}$ as stated in  the following theorem.

\noindent
{\bf Theorem 1.}  The quadratic function $f({\bm t})$ in equation (5) has a minimum in the domain ${\cal D}$, which can be attained  at some finite point ${\bm t}_0\in {\cal D}$.

	Theorem 1 can be proved based on the continuity of function  $f({\bm t})$ in the compact domain ${\cal D}$. This minimization problem is equivalent to a quadratic programming problem using the following Lagrange multiplier for convex programming (\cite{Lee2005}, page 13), for which the existence of the solution is guaranteed by the Khun-Tucker Theorem.
\begin{equation}
\min _{({\bm t},\, {\bm \lambda}, \,\mu)} \left[ f({\bm t}) - (\lambda_1t_1 + \ldots + \lambda_pt_p) +\mu\{(t_1+\ldots+t_p)-1\} \right],
\end{equation}
where ${\bm \lambda}=(\lambda_1, \ldots, \lambda_p)^T$ with $\lambda_1, \ldots, \lambda_p \geq 0$ and $\mu$ is a real number. We also provide the uniqueness of the solution.

\noindent
{\bf Theorem 2.}   The quadratic function $f({\bm t})$ in equation (5) has a unique minimum in the domain ${\cal D}$.

We prove Theorem 2 by contradiction. Assuming that there exist two distinct minima  ${\bm t}_1 \neq {\bm t}_2$ (including local minimum) and further $f({\bm t}_1)\leq f({\bm t}_2)$ without loss of generality, by Jensen's inequality \cite{JensenIneq} one has $f(s{\bm t}_1+(1-s){\bm t}_2) \leq sf({\bm t}_1) + (1-s)f({\bm t}_2) \leq f({\bm t}_2)$ for any real number $0\leq s \leq 1$ with convex function $f({\bm t})$, which implies that ${\bm t}_2$ is not a local or global minimum, unless $f({\bm t}_1)=f(s{\bm t}_1+(1-s){\bm t}_2)= f({\bm t}_2)$, which is impossible for a quadratic function. This contradiction completes the proof.

\noindent
{\bf Algorithm for computing the optimal weights and reference population\hspace{.2em}}
To solve the optimization problem, the following two approaches can be employed. \\
{\bf I. Quadratic programming approach}  The optimization problem in equation (6) leads to an equation system (7) below using the Lagrange multiplier approach by minimizing the objective function, see \cite{Lee2005} for details.
\begin{eqnarray} g ({\bm t}, {\bm \lambda}, \mu) &=& f({\bm t}) - {\bm t}^T {\bm \lambda} + \mu {\bm t}^T {\bm 1} \rule{17em}{0em} \nonumber \\
&=& (M^TN{\bm t}-{\bm R})^T(M^TN{\bm t}-{\bm R}) - {\bm t}^T {\bm \lambda} + \mu ({\bm t}^T {\bm 1}-1) \nonumber
\end{eqnarray}
with parameters ${\bm t}, {\bm \lambda}$ and $\mu$, where $N = ({\bm n}_1, \ldots, {\bm n}_p)$ is a matrix with $p$ column vectors ${\bm n}_1, \ldots, {\bm n}_p$, and  ${\bm 1}$ is a $p$-vector of components 1.
\begin{equation}
\left\{
\begin{array}{rcl} N^TMM^TN {\bm t} &=& N^TM{\bm R}+{\bm \lambda}-\mu {\bm 1} \\
\lambda_it_i &=& 0\;, \;\; i=1,\ldots, p  \\
\sum_{i=1}^p t_i &=& 1\, .
\end{array}\right.
\end{equation}
\noindent
{\bf II. Statistical sampling approach} A statistical sampling method can also find the optimal vector ${\bm t}_0\in {\cal D}$ that minimizes the total deviation $f({\bm t})$. \\
Step (a). Set an initial threshold $\delta = \sum_{i=1}^p b_i^2 ({\bm t}^*)$, i.e. the total deviation with initial value ${\bm t}^* =(1/p, \ldots,1/p)$. \\
Step (b). Take a random sample $(t_1,\ldots, t_{p-1})$ from uniform distribution Unif [0,1] and take the sum. \\
Step (c). If $(t_1+\ldots+ t_{p-1})\leq 1$ set $t_p= 1-(t_1+\ldots+ t_{p-1})$ and go to Step (d). Otherwise, discard the sample and repeat Steps (b) and (c) until the condition  $(t_1+\ldots+ t_{p-1})\leq 1$ is satisfied. \\
Step (d). Check the total deviation $f({\bm t})$ in equation (5). If $f({\bm t}) < \delta$, update the threshold by setting $\delta =  f({\bm t})$. If not, repeat the above steps with a new sample. \\
Step (e). Repeat the above steps (b-d) to achieve a reasonably small total deviation $\delta$. \\
Step (f). Repeat Step (e) to fine-tune the search by shrinking the sampling domain from [0,1] to a small one $[t_i-\delta_{i1}, t_i+\delta_{i2}]$ for $t_i$ ($i=1,\ldots, p-1$) with $\delta_{i2}-\delta_{i1}\rightarrow 0$ and $t_i-\delta_{i1}\geq 0, t_i+\delta_{i1}\leq 1$. This fine-tuning leads to the convergence of ${\bm t}$ by the existence and uniqueness in Theorems 1 and 2.

\noindent
{\bf Comparison of cancer mortality rates by age - standardization methods\hspace{.2em}}
The crude rate was set as the reference for comparison, the cumulative rate and the age-adjusted rates using the US Year 2000 Population Standard and the mean reference population were compared to the crude rate. The relative deviation was calculated and reported as a percentage for each population $i$, and the mean deviation of an age adjusted rate was calculated with equation (8) below as an average over all $p$ populations and used to assess an age adjustment method.
\begin{equation}
\sqrt{\frac{\sum_{i=1}^p ({\bm m}_i^T{\bm n}-{\bm m}_i^T{\bm n}_i)^2 }{p}}
\end{equation}
A large mean deviation indicates inaccurate estimation while a small one indicates accurate estimation.

\noindent
{\bf Software}
SEER*Stat Version 7.1.0 \cite{SeerStat} was used to generate the total US population-based cancer mortality rates. CanQues Version 4.2 \cite{CanQues} was used to calculate disease prevalence. R version 2.13.0 \cite{R} was used for modeling, data analysis, and producing the figures.

\section*{Acknowledgments}
The authors are grateful to Norman Breslow, Robert Rosenberg and Bent Nielsen for their valuable comments and suggestions. We also thank Marla Broadfoot for her editorial assistance. This work is partly supported by NIH grant (5K01CA131259, CA142774 and CA165923). All authors declare no conflict of interest.



\newpage

\begin{table}{\small \begin{center}
\caption{Adjusted Rates Compared to Crude Prostate Cancer Mortality  in Six States  1970--2009$^*$}
\renewcommand{\arraystretch}{.75}
\renewcommand{\tabcolsep}{3pt} 
\begin{tabular}{@{\vrule height 8pt depth1.5pt  width0pt}ccrrrrrrrr}
& & \multicolumn{8}{c}{\underline{\rule{14em}{0em} Year \rule{14em}{0em}}}  \\
State & Method &1970-74&1975-79&1980-84&1985-89&1990-94&1995-99&2000-04&2005-09\\ \hline \hline
CA & Crude & 15.54 & 17.50 & 18.98 & 20.05 & 21.51 & 19.44 & 17.56 & 17.03 \\ \hline
&US2000        & 29.66 & 31.84 & 33.47 & 34.72 & 36.76 & 30.84 & 26.33 & 23.33 \\
&\% Dev\footnote{}
    &  91 &  82 &  76 &  73 &  71 &  59 &  50 &  37 \\ \hline
&MeanRef        & 17.91 & 19.11 & 20.11 & 20.84 & 21.82 & 18.12 & 15.25 & 13.56 \\
&\% Dev    &  15 &  9 &  6 &  4 &  1 & -7 & -13 & -20 \\ \hline
&Cumul    &482.92 &499.10 &511.47 &539.32 &551.92 &440.27 &352.83 &308.69 \\
&\% Dev & 3007 & 2752 & 2594 & 2590 & 2466 & 2165 & 1909 & 1712 \\ \hline  
MA & Crude & 19.21 & 21.25 & 23.64 & 25.13 & 28.80 & 26.69 & 23.59 & 20.78 \\ \hline
&US2000         & 31.31 & 32.68 & 34.57 & 36.01 & 38.87 & 34.01 & 28.66 & 23.17 \\
&\% Dev    &  63 & 54 & 46 & 43 & 35 & 28 & 22 & 12 \\ \hline
&MeanRef         & 23.11 & 24.18 & 25.38 & 26.27 & 28.32 & 24.43 & 20.33 & 16.42 \\
&\% Dev       & 20 & 14  & 7 &  5 & -2 & -8 & -14 & -21 \\ \hline
&Cumul    &499.05 &527.05 &529.31 &528.14 &567.52 &473.18 &359.01 &275.71 \\
&\% Dev & 2498 & 2381 & 2139 & 2002 & 1871 & 1673 & 1422 & 1227 \\ \hline  
MI & Crude & 16.63 & 18.51 & 20.82 & 23.84 & 27.26 & 25.14 & 21.44 & 18.85 \\ \hline
&US2000         & 31.68 & 33.91 & 34.99 & 37.33 & 40.91 & 36.09 & 28.71 & 22.45 \\
&\% Dev     &  90 &  83 &  68 &  57 &  50 &  44 &  34 &  19 \\ \hline
&MeanRef         & 21.06 & 22.27 & 22.90 & 24.38 & 26.15 & 22.64 & 17.88 & 14.04 \\
&\% Dev     &  27 &  20 &  10 &  2 & -4 & -10 & -17 & -25 \\ \hline
&Cumul     &539.70 &549.20 &569.68 &585.61 &616.35 &501.73 &384.85 &289.05 \\
&\% Dev & 3145 & 2868 & 2637 & 2357 & 2161 & 1896 & 1695 & 1434 \\ \hline  
MO & Crude & 20.93 & 22.87 & 23.38 & 26.21 & 28.96 & 26.11 & 21.01 & 19.99 \\ \hline
&US2000         & 29.37 & 31.78 & 31.18 & 33.84 & 36.61 & 32.94 & 26.10 & 23.09 \\
&\% Dev     &  40 &  39 &  33 &  29 &  26 &  26 &  24 &  15 \\ \hline
&MeanRef         & 23.30 & 24.85 & 24.30 & 26.39 & 28.38 & 25.05 & 19.51 & 17.49 \\
&\% Dev    &  11 &  9 &  4 &  1 & -2 & -4 & -7 & -13 \\ \hline
&Cumul    &495.14 &502.12 &481.57 &524.52 &540.68 &459.26 &325.60 &308.95 \\
&\% Dev & 2265 & 2095 & 1959 & 1901 & 1767 & 1659 & 1449 & 1445 \\ \hline  
NJ & Crude & 17.53 & 20.38 & 23.09 & 25.70 & 29.68 & 26.98 & 22.43 & 19.08 \\ \hline
&US2000         & 31.36 & 33.42 & 35.48 & 36.95 & 40.80 & 35.19 & 28.13 & 22.30 \\
&\% Deviate     &  79 &  64 &  54 &  44 &  37 &  30 &  25 &  17 \\ \hline
&MeanRef         & 22.28 & 23.75 & 24.91 & 26.11 & 28.46 & 24.37 & 19.19 & 15.28 \\
&\% Dev     &  27 &  17 &  8 &  2 & -4 & -10 & -14 & -20 \\ \hline
&Cumul     &524.23 &542.64 &546.12 &574.13 &604.50 &510.12 &373.19 &302.53 \\
&\% Dev & 2891 & 2563 & 2265 & 2134 & 1937 & 1791 & 1563 & 1485 \\ \hline  
NY & Crude & 18.15 & 20.57 & 22.94 & 24.58 & 27.23 & 24.83 & 22.12 & 19.30 \\ \hline
&US2000         & 28.61 & 31.12 & 33.13 & 34.42 & 37.80 & 33.27 & 28.02 & 22.49 \\
&\% Dev     &  58 &  51 &  44 &  40 &  39 &  34 &  27 &  16 \\ \hline
&MeanRef         & 21.22 & 22.74 & 24.02 & 25.16 & 27.13 & 23.66 & 19.71 & 15.83 \\
&\% Dev     &  17 &  11 &  5 &  2 &  0 & -5 & -11 & -18 \\ \hline
&Cumul     &486.59 &504.17 &529.21 &548.42 &572.84 &489.36 &386.83 &310.59 \\
&\% Dev & 2581 & 2351 & 2207 & 2131 & 2003 & 1871 & 1649 & 1509 \\ \hline   \hline
\rule{0em}{.01em} \\
  \multicolumn{8}{l}{$^*$ Unit of mortality rate is per $10^5$  person-year.}\\
\multicolumn{8}{l}{$^1$ Percentage of deviation from the crude rate.}
\end{tabular} \end{center} }
\end{table}

\renewcommand{\baselinestretch}{1.6}

\begin{table} { \small \begin{center}
\caption{Expected Number of Deaths by Summary Rates Compared to Observed Number of Deaths}
\renewcommand{\arraystretch}{.75}
\renewcommand{\tabcolsep}{3pt} 
\begin{tabular}{@{\vrule height 8pt depth1.5pt  width0pt}ccrrrrrrrr}
& & \multicolumn{8}{c}{\underline{\rule{13em}{0em} Year \rule{13em}{0em}}}  \\
State & Method &1970-74&1975-79&1980-84&1985-89&1990-94&1995-99&2000-04&2005-09\\ \hline  \hline
CA & Obs & 7760 &  9503 & 11457 & 13657 & 16276 & 15530 & 15000 & 15169 \\
   & Crude  & 7760 & 9503 & 11457 & 13657 & 16276 & 15530 & 15000 & 15169 \\ \hline
&US2000        & 14807 & 17290 & 20203 & 23649 & 27809 & 24639 & 22492 & 20772  \\
&\% Dev\footnote{}    &  91 &  82 &  76 &  73 &  71 &  59 &  50 &  37 \\ \hline
&MeanRef        & 8940 & 10380 & 12136 & 14192 & 16509 & 14474 & 13030 & 12075 \\
&\% Dev    &  15 &  9 &  6 &  4 &  1 & -7 & -13 & -20 \\ \hline
MA & Obs &  2600 & 2872 & 3215 & 3521 & 4119 & 3945  &3597 & 3199 \\
&Crude &    2600 & 2872 & 3215 & 3521 & 4119 & 3945  &3597 & 3199 \\ \hline
&US2000     &  4238  & 4417 & 4701 & 5046 & 5561 & 5028  &4370 & 3567 \\
&\% Dev    &  63 & 54 & 46 & 43 & 35 & 28 & 22 & 12 \\ \hline
&MeanRef         & 3129 & 3268 & 3452 & 3682 & 4050 & 3611 & 3100 & 2528 \\
&\% Dev       & 20 & 14  & 7 &  5 & -2 & -9 & -14 & -21 \\ \hline
MI & Obs & 3610 &  4078 &  4553 &  5223 &  6180  & 5938 & 5194 & 4561 \\
& Crude & 3610 &  4078 &  4553 &  5223 &  6180  & 5938 & 5194 & 4561 \\ \hline
&US2000     & 6876 &  7473  & 7654  & 8181 &  9274  & 8525 & 6956 & 5433 \\
&\% Dev     &  90 &  83 &  68 &  57 &  50 &  44 &  34 &  19 \\ \hline
&MeanRef         & 4570 &  4908 &  5009 &  5341 &  5927  & 5347 & 4332 & 3398 \\
&\% Dev     &  27 &  20 &  10 &  2 & -4 & -10 & -17 & -25 \\ \hline
MO & Obs & 2357 & 2627 & 2732 & 3135 & 3595 & 3418 & 2865 & 2835 \\
& Crude & 2357 & 2627 & 2732 & 3135 & 3595 & 3418 & 2865 & 2835 \\ \hline
&US2000         & 3307 & 3650 & 3643 & 4047 & 4544 & 4311 & 3558 & 3273 \\
&\% Dev     &  40 &  39 &  33 &  29 &  26 &  26 &  24 &  15 \\ \hline
&MeanRef         & 2623 & 2854 & 2839 & 3156 & 3523 & 3280 & 2661 & 2481 \\
&\% Dev    &  11 &  9 &  4 &  1 & -2 & -4 & -7 & -13 \\ \hline
NJ & Obs & 3042 & 3553 & 4068 &  4674  & 5570  & 5298 & 4589 & 3983 \\
& Crude & 3042 & 3553 & 4068 &  4674  & 5570  & 5298 & 4589 & 3983 \\ \hline
&US2000         & 5442 & 5826 & 6251 &  6721  & 7657 &  6912 & 5754 & 4655 \\
&\% Dev     &  79 &  64 &  54 &  44 &  37 &  30 &  25 &  17 \\ \hline
&MeanRef         & 3867 & 4141 & 4388 &  4750 &  5342 &  4786 & 3926 & 3190 \\
&\% Dev     &  27 &  17 &  8 &  2 & -4 & -10 & -14 & -20 \\ \hline
NY & Obs & 7791  & 8616  & 9476 & 10343 & 11739 & 11016 & 10061 &  8832 \\
 & Crude & 7791  & 8616  & 9476 & 10343 & 11739 & 11016 & 10061 &  8832 \\ \hline
&US2000      & 12282 & 13036 & 13688 & 14482&  16296 & 14761 & 12746 & 10288  \\
&\% Dev     &  58 &  51 &  44 &  40 &  39 &  34 &  27 &  16 \\ \hline
&MeanRef         & 9111 &  9526 &  9922 & 10584 & 11694 & 10497 &  8963 &  7243 \\
&\% Dev     &  17 &  11 &  5 &  2 &  0 & -5 & -11 & -18 \\ \hline  \hline
\rule{0em}{.01em} \\
\multicolumn{8}{l}{$^2$ Percentage of deviation from the observed number.}
\end{tabular} \end{center} }
\end{table}

\renewcommand{\baselinestretch}{2}

\begin{table} {\small \begin{center}
\caption{Comparison of Age-adjusted Case Fatality Rate to Crude Rate in US in 2008$^*$}
\renewcommand{\arraystretch}{.75}
\renewcommand{\tabcolsep}{3pt} 
\begin{tabular}{@{\vrule height 9pt depth1.5pt  width0pt}ccrrrrrr}
   & Mean & \multicolumn{6}{c}{\underline{\rule{10em}{0em} Cancer Site \rule{10em}{0em}}} \\
Rate & Deviation$^b$ &   Breast & Cervix & Prostate & Lung & Leukemia & Colon-Rect  \\ \hline \hline
  Crude & 0 &  1380.6&  1170.5&  1088.3 & 19771.7& 3705.3& 3201.5 \\  \hline
  US2000 & 4305.1  & 941.7&  683.9&  243.4& 9384.5& 2265.3& 2900.3 \\
  \% Dev$^a$&  & -31.79 &   -41.57 &  -77.63  & -52.54  & -38.86  & -9.41 \\ \hline
 MeanRef & 403.4 & 1394.1&  1295.7& 845.2& 19344.1& 4538.4& 3358.1 \\
  \% Dev$^a$& &  0.98 &  10.70 & -22.34 & -2.16 & 22.48 & 4.89  \\ \hline
  Cumul  & 53351.3 & 10867.8  &  9789.3 &   4175.9 & 145146.2 &  28241.8 &  27358.2 \\
  \% Dev&   & 687.19 &    736.36 &    283.71  &   634.11  &   662.19  &   754.55 \\ \hline \hline
  \rule{0em}{.5em} \\
  \multicolumn{8}{l}{$^*$ Unit of case fatality rate is per $10^5$  person-year.}\\
  \multicolumn{8}{l}{$^a$ Negative deviation indicates underestimation.} \\
  \multicolumn{8}{l}{$^b$ Mean deviation is calculated as the square-root of the average of the squared} \\
   \multicolumn{8}{l}{ deviation over all six cancer sites, see equation (8) in Methods.}
\end{tabular} \end{center} }
\end{table}



\newpage
\clearpage
\section*{SUPPLEMENTARY MATERIALS}
\setcounter{page}{1}

\section*{Data}

\subsection*{Generating Cancer Mortality  and Case Fatality Rates}


We generate prostate cancer mortality rate of each state with the US SEER 9 registration data using the {\it Mortality All COD - Aggregated with State, Total US (1969-2009) $<$Katrina/Rita Population Adjustment$>$} with the state specified to be one of the six states  and gender specified to be ``Male" and Site and Morphology Cause of Death Record to be ``Prostate". Eight 5 year periods (1970--74, $\dots$, 2005--09) were specified to generate the age--specific cancer mortality rate for the eight periods. By default, the numbers of cases less than 10 were set to 0 with a corresponding mortality rate 0. The newborn group age 0 was excluded for data analysis and the remaining 18 age groups (1--4, 5--9, $\ldots$, 80-84, and 85+) were used for data analysis.


We generate the cancer mortality rate with the US SEER 9 registry data using the incidence-based mortality (IBM) database ({\it 31}), the {\it Incidence-Based Mortality - SEER 9 Regs Research Data, Nov 2011 Sub Vintage 2009 Pops (1973-2009) $<$Katrina/Rita Population Adjustment$>$}. Following the IBM instruction ({\it 31}), we specify the maximum number of months of the ``Survival time recode'' in the database to ensure that all cancer records in the database are included in calculating the age-specific mortality rate for the year 2008. Also generated by the SEER*Stat software are the numbers of deaths reported in 2008 by age group and the numbers of people in the general public (including healthy people). Similar to the above, we exclude the age 0 group and thus have the mortality rates, the numbers of deaths, and the population exposures  in 18 age groups (age 1-4, 5-9, $\ldots, 80$ and $85+$).  Note that although SEER*Stat is used to estimate mortality rate based on the total population in the database, it does not provide patient exposure-based case fatality rate ({\it 25}). We further generate the prevalence of each cancer as of January 1, 2009  by the same 18 age groups using CanQues, and calculate the case fatality rate as follows.

Let $m_p$ denote the mortality rate based on the general population (including the healthy people), $m_c$  the case fatality rate based on the cancer patient exposure to death from the disease, $D$  the number of deaths from the disease,  $E$  the cancer patient exposure, $Pop$  the total population, and $Prev$  the prevalence of the disease in the general population. It can be seen that
$$ m_c = \frac{D}{E} = \frac{D}{Pop \times Prev} = \frac{m_p}{Prev}, \eqno{[9]} $$
i.e., each age-specific case fatality rate $m_c$ can be calculated with the age-specific mortality rate  $m_p$ divided by  the age-specific  prevalence of the disease $Prev$.  Note that the following two rules apply in calculating the mortality rates: \\
1) If the age-specific prevalence $Prev= 0$, the mortality rate $m_c$ is set to be 0. \\
2) To ensure the stability of mortality rate based on a small number of cases, the age-specific case fatality rate $m_c$ is set to be 0 if the age-specific number of deaths is small ($D < 5$).

Table S1 displays the person-year exposure in 2008 by cancer site and age group and the US Year 2000 Population Standard by age group. Table S2 displays the case fatality rates in 2008 by age group of six cancer sites: female breast, cervix, prostate, lung, leukemia, and colon-rectum.
Tables S3--S14 in the Supplementary Materials display the mortality rate and population exposure during 1970--2009 of the six states. The newborn group of age 0 is excluded from all tables  as they are not considered for cancer mortality in this study.

\clearpage
\newpage

\begin{center}
\begin{table} 
\rule{0em}{1em}
\begin{center} {Table S1. Distribution of US Cancer Patient of Six Sites and US Year 2000 Population Standard$^1$} \\
\renewcommand{\arraystretch}{.8}
\begin{tabular}{crrrrrrrrr}
Age & \multicolumn{6}{c}{\underline{\rule{13em}{0em} Cancer Site \rule{13em}{0em}} } & US 2000  \\ 
(year) & Female Breast & Cervix & Prostate & Lung & Leukemia & Colon-Rectum  &  ($\times 1000$)\\ \hline \hline
1-4 &     0.0&    0.0&      0.0&    0.0&    288.4&        0.0 & 15191.6\\ 
5-9 &     0.0&    0.0&      0.0&    0.0&    798.1&        0.0 & 19919.8 \\ 
10-14 &     2.0&    2.9&      0.0&    2.1&    963.7&        0.0 & 20056.8 \\ 
15-19 &     6.0&    3.0&      0.0&    9.3&   1169.0&        0.0 & 19819.5 \\ 
20-24 &    30.3&   37.9&      0.0&   23.5&   1016.5&       51.0 & 18257.2\\ 
25-29 &   209.5&  147.2&      0.0&   54.5&   1015.0&      167.8 & 17722.0 \\ 
30-34 &   825.1&  457.9&      0.0&   78.0&    898.8&      335.3 & 19511.4 \\ 
35-39 &  2665.1& 1110.3&     18.6&  152.1&    893.8&      696.6 & 22180.0 \\ 
40-44 &  6535.6& 1681.6&    279.9&  345.9&    802.9&     1574.9 & 22479.2 \\ 
45-49 & 13663.2& 2315.5&   1534.3&  916.9&   1071.2&     2900.7 & 19805.8 \\ 
50-54 & 21312.3& 2683.2&   5756.7& 1926.7&   1332.3&     5505.3 & 17224.4 \\ 
55-59 & 27213.8& 2763.8&  13937.6& 2840.2&   1699.5&     7808.3 & 13307.2 \\ 
60-64 & 31922.1& 2274.7&  25540.1& 4160.4&   2044.7&     9616.8 & 10654.2 \\ 
65-69 & 30378.7& 1690.6&  33205.4& 5175.3&   2079.9&    10896.1 & 9409.9 \\ 
70-74 & 26981.9& 1256.4&  36867.2& 5293.5&   2028.9&    12527.7 & 8725.6 \\ 
75-79 & 25525.1&  961.1&  38201.6& 5244.7&   2009.4&    14127.7 & 7414.6 \\ 
80-84 & 23685.1&  617.2&  33037.0& 4189.4&   1815.1&    14773.0 & 4900.2 \\ 
85+ & 25249.0&  536.8&  26454.0& 2766.4&   1687.5&    18098.4 & 4259.2 \\ 
 \hline \hline
\end{tabular}
\end{center}
\rule{0em}{2em}
$^1$Fraction person-year exposure is due to the conversion from total population by disease prevalence.
\end{table}
\end{center}

\begin{center}
\begin{table}
\rule{0em}{1em}
\begin{center}{Table S2. US Age-specific Case Fatality Rate ($10^{-5}$ person-year) of six Cancer Sites  in 2008} \\
\renewcommand{\arraystretch}{.8}
\begin{tabular}{crrrrrrrr}
Age & \multicolumn{6}{c}{\underline{\rule{13em}{0em} Cancer Site \rule{13em}{0em}} } \\ (year) & 
  Female Breast & Cervix & Prostate & Lung & Leukemia & Colon-Rectum  \\ \hline \hline
1-4 &    0.0&    0.0&      0.0&     0.0&   2424.6&        0.0 \\
5-9 &    0.0&    0.0&      0.0&     0.0&    876.5&        0.0 \\
10-14 &    0.0&    0.0&      0.0&     0.0&    623.0&        0.0 \\
15-19 &    0.0&    0.0&      0.0&     0.0&    599.3&        0.0 \\
20-24 &    0.0&    0.0&      0.0&     0.0&   1474.9&        0.0 \\
25-29 &    0.0&    0.0&      0.0&     0.0&   1380.2&     7750.0 \\
30-34 & 1575.9& 1092.6&      0.0& 10250.0&   1780.9&     4470.9 \\
35-39 & 2363.8& 1260.8&      0.0& 10527.0&   1119.5&     6174.0 \\
40-44 & 1713.7&  832.8&      0.0& 11857.1&   2615.4&     4508.5 \\
45-49 & 1690.6&  906.8&    456.5& 18213.3&   2148.1&     4171.5 \\
50-54 & 1398.2& 1304.4&    330.1& 18322.3&   2552.3&     3306.0 \\
55-59 & 1355.9& 1085.5&    466.4& 19470.5&   3294.6&     3393.8 \\
60-64 & 1181.0&  967.3&    536.4& 19132.5&   3618.8&     3098.7 \\
65-69 & 1168.6& 1656.3&    605.3& 17796.1&   3269.5&     3184.6 \\
70-74 & 1145.2&  955.0&    783.9& 19552.3&   5766.6&     2785.8 \\
75-79 & 1214.5& 2081.2&    997.3& 20802.1&   4976.5&     2909.2 \\
80-84 & 1397.5& 1458.2&   1440.8& 21292.0&   7823.0&     2931.0 \\
85+ & 1952.6& 1303.7&   2884.2& 24833.7&   9362.8&     3447.8
 \\  \hline \hline
\end{tabular}
\end{center}
\end{table}
\end{center}

\begin{table}
\begin{center}{Table S3. Prostate Cancer Mortality Rate ($10^{-5}$) during 1970 -- 2009 in California}
\begin{tabular}{rrrrrrrrr}
Age &1970-74&1975-79&1980-84&1985-89&1990-94&1995-99&2000-04&2005-09\\ \hline \hline
1-4   &0.00   &0.00   &0.00   &0.00   &0.00   &0.00   &0.00   &0.00\\ \hline
5-9   &0.00   &0.00   &0.00   &0.00   &0.00   &0.00   &0.00   &0.00\\ \hline
10-14   &0.00   &0.00   &0.00   &0.00   &0.00   &0.00   &0.00   &0.00\\ \hline
15-19   &0.00   &0.00   &0.00   &0.00   &0.00   &0.00   &0.00   &0.00\\ \hline
20-24   &0.00   &0.00   &0.00   &0.00   &0.00   &0.00   &0.00   &0.00\\ \hline
25-29   &0.00   &0.00   &0.00   &0.00   &0.00   &0.00   &0.00   &0.00\\ \hline
30-34   &0.00   &0.00   &0.00   &0.00   &0.00   &0.00   &0.00   &0.00\\ \hline
35-39   &0.00   &0.00   &0.00   &0.00   &0.00   &0.00   &0.00   &0.00\\ \hline
40-44   &0.00   &0.00   &0.00   &0.22   &0.20   &0.00   &0.30   &0.18\\ \hline
45-49   &0.67   &0.79   &0.78   &0.89   &0.71   &0.80   &0.75   &0.54\\ \hline
50-54   &3.61   &3.73   &4.48   &3.99   &3.57   &3.10   &2.95   &2.69\\ \hline
55-59  &11.07  &12.39  &13.31  &12.60  &13.12  &10.99   &8.11   &8.68\\ \hline
60-64  &30.86  &31.37  &35.91  &36.06  &35.76  &28.59  &23.70  &21.08\\ \hline
65-69  &64.99  &69.14  &74.57  &76.33  &77.50  &62.18  &46.59  &43.93\\ \hline
70-74 &134.76 &142.94 &138.91 &151.10 &150.68 &123.16  &91.81  &79.66\\ \hline
75-79 &236.94 &238.73 &243.51 &258.12 &270.40 &211.45 &178.62 &151.92\\ \hline
80-84 &362.30 &388.02 &413.94 &425.28 &444.56 &355.77 &315.44 &280.77\\ \hline
85+   &507.62 &580.92 &625.84 &639.94 &726.02 &672.19 &607.72 &541.34\\ \hline \hline
\end{tabular}
\end{center}
\end{table}

\begin{table}
\begin{center}{Table S4. Population Proportions during 1970 -- 2009 in California}
\begin{tabular}{rrrrrrrrr}
Age &1970-74&1975-79&1980-84&1985-89&1990-94&1995-99&2000-04&2005-09\\ \hline \hline
1-4   &0.066 &0.058 &0.063 &0.066 &0.073 &0.069 &0.060 &0.058\\ \hline
5-9   &0.092 &0.079 &0.071 &0.075 &0.079 &0.085 &0.078 &0.070\\ \hline
10-14 &0.101 &0.089 &0.077 &0.068 &0.073 &0.077 &0.081 &0.077\\ \hline
15-19 &0.099 &0.099 &0.087 &0.079 &0.070 &0.073 &0.077 &0.081\\ \hline
20-24 &0.096 &0.103 &0.105 &0.097 &0.085 &0.072 &0.078 &0.078\\ \hline
25-29 &0.082 &0.095 &0.105 &0.106 &0.096 &0.085 &0.076 &0.077\\ \hline
30-34 &0.067 &0.080 &0.093 &0.099 &0.099 &0.090 &0.081 &0.072\\ \hline
35-39 &0.058 &0.062 &0.073 &0.084 &0.088 &0.089 &0.082 &0.075\\ \hline
40-44 &0.059 &0.055 &0.057 &0.066 &0.074 &0.078 &0.081 &0.076\\ \hline
45-49 &0.060 &0.054 &0.049 &0.051 &0.058 &0.066 &0.072 &0.075\\ \hline
50-54 &0.056 &0.054 &0.047 &0.043 &0.045 &0.053 &0.061 &0.067\\ \hline
55-59 &0.047 &0.049 &0.047 &0.040 &0.037 &0.039 &0.047 &0.056\\ \hline
60-64 &0.038 &0.040 &0.041 &0.038 &0.033 &0.032 &0.035 &0.042\\ \hline
65-69 &0.030 &0.032 &0.033 &0.033 &0.031 &0.029 &0.028 &0.030\\ \hline
70-74 &0.021 &0.023 &0.024 &0.024 &0.025 &0.025 &0.024 &0.023\\ \hline
75-79 &0.014 &0.015 &0.016 &0.016 &0.017 &0.019 &0.019 &0.019\\ \hline
80-84 &0.008 &0.008 &0.009 &0.009 &0.010 &0.011 &0.013 &0.013\\ \hline
85+   &0.005 &0.005 &0.006 &0.006 &0.006 &0.007 &0.008 &0.010\\ \hline \hline
\end{tabular}
\end{center}
\end{table}

\begin{table}
\begin{center}{Table S5. Prostate Cancer Mortality Rate ($10^{-5}$) during 1970 -- 2009 in Massachusetts}
\begin{tabular}{rrrrrrrrr}
Age &1970-74&1975-79&1980-84&1985-89&1990-94&1995-99&2000-04&2005-09\\ \hline \hline
1-4   &0.00   &0.00   &0.00   &0.00   &0.00   &0.00   &0.00   &0.00\\ \hline
5-9   &0.00   &0.00   &0.00   &0.00   &0.00   &0.00   &0.00   &0.00\\ \hline
10-14   &0.00   &0.00   &0.00   &0.00   &0.00   &0.00   &0.00   &0.00\\ \hline
15-19   &0.00   &0.00   &0.00   &0.00   &0.00   &0.00   &0.00   &0.00\\ \hline
20-24   &0.00   &0.00   &0.00   &0.00   &0.00   &0.00   &0.00   &0.00\\ \hline
25-29   &0.00   &0.00   &0.00   &0.00   &0.00   &0.00   &0.00   &0.00\\ \hline
30-34   &0.00   &0.00   &0.00   &0.00   &0.00   &0.00   &0.00   &0.00\\ \hline
35-39   &0.00   &0.00   &0.00   &0.00   &0.00   &0.00   &0.00   &0.00\\ \hline
40-44   &0.00   &0.00   &0.00   &0.00   &0.00   &0.00   &0.00   &0.00\\ \hline
45-49   &0.00   &0.00   &1.68   &0.00   &0.00   &0.00   &0.00   &1.20\\ \hline
50-54   &4.31   &4.72   &3.46   &3.64   &4.12   &3.69   &3.17   &3.34\\ \hline
55-59   &9.80   &9.48  &12.45  &11.42  &13.51  &11.57   &8.68   &6.38\\ \hline
60-64  &26.45  &27.16  &33.75  &35.30  &37.39  &22.79  &22.52  &16.54\\ \hline
65-69  &68.11  &74.38  &73.67  &70.23  &76.88  &57.33  &48.72  &39.35\\ \hline
70-74 &135.62 &144.99 &142.55 &154.44 &164.39 &131.34  &91.42  &72.20\\ \hline
75-79 &254.76 &266.32 &261.75 &253.10 &271.25 &246.46 &184.51 &136.69\\ \hline
80-84 &419.33 &437.60 &405.24 &433.59 &444.32 &432.50 &333.59 &289.91\\ \hline
85+   &522.41 &532.91 &676.40 &740.08 &829.64 &732.48 &726.27 &586.71\\ \hline \hline
\end{tabular}
\end{center}
\end{table}

\begin{table}
\begin{center}{Table S6. Population Proportions during 1970 -- 2009 in Massachusetts}
\begin{tabular}{rrrrrrrrr}
Age &1970-74&1975-79&1980-84&1985-89&1990-94&1995-99&2000-04&2005-09\\ \hline \hline
1-4   &0.067 &0.054 &0.052 &0.057 &0.061 &0.057 &0.052 &0.049\\ \hline
5-9   &0.095 &0.080 &0.065 &0.065 &0.071 &0.076 &0.069 &0.064\\ \hline
10-14 &0.106 &0.096 &0.081 &0.065 &0.065 &0.070 &0.074 &0.069\\ \hline
15-19 &0.100 &0.104 &0.095 &0.082 &0.065 &0.066 &0.072 &0.077\\ \hline
20-24 &0.087 &0.096 &0.103 &0.096 &0.085 &0.066 &0.068 &0.073\\ \hline
25-29 &0.074 &0.086 &0.095 &0.102 &0.093 &0.082 &0.068 &0.068\\ \hline
30-34 &0.059 &0.073 &0.084 &0.091 &0.094 &0.087 &0.076 &0.064\\ \hline
35-39 &0.053 &0.057 &0.069 &0.081 &0.086 &0.089 &0.083 &0.072\\ \hline
40-44 &0.057 &0.051 &0.054 &0.067 &0.076 &0.081 &0.085 &0.079\\ \hline
45-49 &0.059 &0.054 &0.048 &0.052 &0.062 &0.072 &0.077 &0.081\\ \hline
50-54 &0.058 &0.056 &0.051 &0.045 &0.048 &0.059 &0.068 &0.074\\ \hline
55-59 &0.051 &0.053 &0.052 &0.046 &0.041 &0.044 &0.054 &0.064\\ \hline
60-64 &0.043 &0.045 &0.047 &0.045 &0.041 &0.037 &0.040 &0.049\\ \hline
65-60 &0.033 &0.036 &0.038 &0.039 &0.038 &0.035 &0.032 &0.035\\ \hline
70-74 &0.025 &0.026 &0.028 &0.029 &0.031 &0.031 &0.029 &0.027\\ \hline
75-79 &0.017 &0.017 &0.018 &0.020 &0.022 &0.024 &0.024 &0.023\\ \hline
80-84 &0.010 &0.011 &0.011 &0.011 &0.013 &0.015 &0.016 &0.017\\ \hline
85+   &0.006 &0.007 &0.008 &0.008 &0.008 &0.009 &0.011 &0.013\\ \hline \hline
\end{tabular}
\end{center}
\end{table}

\begin{table}
\begin{center}{Table S7. Prostate Cancer Mortality Rate ($10^{-5}$) during 1970 -- 2009 in Michigan}
\begin{tabular}{rrrrrrrrr}
Age &1970-74&1975-79&1980-84&1985-89&1990-94&1995-99&2000-04&2005-09\\ \hline \hline
1-4   &0.00   &0.00   &0.00   &0.00   &0.00   &0.00   &0.00   &0.00\\ \hline
5-9   &0.00   &0.00   &0.00   &0.00   &0.00   &0.00   &0.00   &0.00\\ \hline
10-14   &0.00   &0.00   &0.00   &0.00   &0.00   &0.00   &0.00   &0.00\\ \hline
15-19   &0.00   &0.00   &0.00   &0.00   &0.00   &0.00   &0.00   &0.00\\ \hline
20-24   &0.00   &0.00   &0.00   &0.00   &0.00   &0.00   &0.00   &0.00\\ \hline
25-29   &0.00   &0.00   &0.00   &0.00   &0.00   &0.00   &0.00   &0.00\\ \hline
30-34   &0.00   &0.00   &0.00   &0.00   &0.00   &0.00   &0.00   &0.00\\ \hline
35-39   &0.00   &0.00   &0.00   &0.00   &0.00   &0.00   &0.00   &0.00\\ \hline
40-44   &0.00   &0.00   &0.00   &0.00   &0.00   &0.00   &0.00   &0.00\\ \hline
45-49   &1.26   &1.11   &1.34   &1.04   &0.00   &1.11   &1.00   &0.73\\ \hline
50-54   &3.74   &2.73   &3.55   &4.47   &3.82   &2.76   &3.13   &3.46\\ \hline
55-59  &14.04  &14.17  &13.14  &15.85  &12.43  &11.77   &9.82   &8.56\\ \hline
60-64  &31.78  &34.80  &36.82  &38.11  &38.41  &29.30  &21.38  &20.18\\ \hline
65-69  &79.88  &81.33  &81.75  &82.80  &84.97  &66.60  &50.59  &40.86\\ \hline
70-74 &155.20 &158.34 &158.39 &165.75 &168.89 &135.68 &102.56  &81.66\\ \hline
75-79 &253.81 &256.72 &274.69 &277.57 &307.82 &254.50 &196.35 &133.60\\ \hline
80-84 &398.95 &425.01 &419.97 &474.40 &498.26 &432.96 &338.74 &252.75\\ \hline
85+   &474.85 &569.07 &605.01 &654.68 &808.06 &802.12 &670.45 &551.83\\ \hline \hline
\end{tabular}
\end{center}
\end{table}

\begin{table}
\begin{center}{Table S8. Population Proportions during 1970 -- 2009 in Michigan} \begin{tabular}{rrrrrrrrr}
Age &1970-74&1975-79&1980-84&1985-89&1990-94&1995-99&2000-04&2005-09\\ \hline \hline
1-4   &0.074 &0.064 &0.063 &0.063 &0.065 &0.060 &0.056 &0.053\\ \hline
5-9   &0.102 &0.090 &0.079 &0.080 &0.079 &0.081 &0.075 &0.070\\ \hline
10-14 &0.114 &0.101 &0.090 &0.077 &0.078 &0.078 &0.081 &0.076\\ \hline
15-19 &0.106 &0.107 &0.096 &0.086 &0.076 &0.077 &0.077 &0.080\\ \hline
20-24 &0.084 &0.095 &0.097 &0.086 &0.077 &0.068 &0.071 &0.071\\ \hline
25-29 &0.074 &0.086 &0.092 &0.091 &0.079 &0.072 &0.063 &0.063\\ \hline
30-34 &0.060 &0.072 &0.082 &0.088 &0.088 &0.079 &0.072 &0.061\\ \hline
35-39 &0.053 &0.057 &0.067 &0.079 &0.086 &0.086 &0.076 &0.069\\ \hline
40-44 &0.057 &0.051 &0.054 &0.065 &0.077 &0.082 &0.082 &0.074\\ \hline
45-49 &0.058 &0.053 &0.048 &0.053 &0.062 &0.072 &0.078 &0.079\\ \hline
50-54 &0.055 &0.055 &0.050 &0.046 &0.050 &0.058 &0.068 &0.075\\ \hline
55-59 &0.048 &0.049 &0.050 &0.046 &0.042 &0.045 &0.054 &0.065\\ \hline
60-64 &0.039 &0.040 &0.043 &0.044 &0.041 &0.037 &0.041 &0.050\\ \hline
65-69 &0.029 &0.030 &0.034 &0.037 &0.037 &0.034 &0.032 &0.037\\ \hline
70-74 &0.021 &0.022 &0.024 &0.027 &0.028 &0.029 &0.028 &0.027\\ \hline
75-79 &0.014 &0.014 &0.016 &0.018 &0.020 &0.021 &0.023 &0.022\\ \hline
80-84 &0.008 &0.008 &0.009 &0.010 &0.011 &0.012 &0.014 &0.016\\ \hline
85+   &0.004 &0.005 &0.006 &0.006 &0.007 &0.008 &0.009 &0.011\\ \hline \hline
\end{tabular}
\end{center}
\end{table}

\begin{table}
\begin{center}{Table S9. Prostate Cancer Mortality Rate ($10^{-5}$) during 1970 -- 2009 in Missouri}
\begin{tabular}{rrrrrrrrr}
Age &1970-74&1975-79&1980-84&1985-89&1990-94&1995-99&2000-04&2005-09\\ \hline \hline
1-4   &0.00   &0.00   &0.00   &0.00   &0.00   &0.00   &0.00   &0.00\\ \hline
5-9   &0.00   &0.00   &0.00   &0.00   &0.00   &0.00   &0.00   &0.00\\ \hline
10-14   &0.00   &0.00   &0.00   &0.00   &0.00   &0.00   &0.00   &0.00\\ \hline
15-19   &0.00   &0.00   &0.00   &0.00   &0.00   &0.00   &0.00   &0.00\\ \hline
20-24   &0.00   &0.00   &0.00   &0.00   &0.00   &0.00   &0.00   &0.00\\ \hline
25-29   &0.00   &0.00   &0.00   &0.00   &0.00   &0.00   &0.00   &0.00\\ \hline
30-34   &0.00   &0.00   &0.00   &0.00   &0.00   &0.00   &0.00   &0.00\\ \hline
35-39   &0.00   &0.00   &0.00   &0.00   &0.00   &0.00   &0.00   &0.00\\ \hline
40-44   &0.00   &0.00   &0.00   &0.00   &0.00   &0.00   &0.00   &0.00\\ \hline
45-49   &0.00   &0.00   &0.00   &0.00   &1.48   &1.33   &1.07   &0.99\\ \hline
50-54   &4.06   &4.34   &3.75   &4.77   &4.01   &5.21   &3.78   &2.64\\ \hline
55-59  &11.85  &10.95  &10.40  &13.59  &15.16  &10.65   &6.96   &8.59\\ \hline
60-64  &30.92  &29.08  &34.20  &36.73  &34.97  &26.99  &21.38  &22.32\\ \hline
65-69  &73.89  &74.38  &68.84  &73.84  &74.84  &62.27  &46.60  &41.98\\ \hline
70-74 &133.31 &131.75 &130.50 &135.31 &151.68 &130.53  &87.08  &83.02\\ \hline
75-79 &241.11 &251.63 &233.88 &260.29 &258.53 &222.28 &158.73 &149.41\\ \hline
80-84 &341.65 &393.60 &373.19 &402.01 &463.89 &382.68 &316.66 &272.11\\ \hline
85+   &487.69 &571.25 &593.84 &642.35 &709.94 &734.11 &642.25 &534.02\\ \hline \hline
\end{tabular}
\end{center}
\end{table}

\begin{table}
\begin{center}{Table S10. Population Proportions during 1970 -- 2009 in Missouri}
\begin{tabular}{rrrrrrrrr}
Age &1970-74&1975-79&1980-84&1985-89&1990-94&1995-99&2000-04&2005-09\\ \hline \hline
1-4   &0.067 &0.061 &0.064 &0.064 &0.062 &0.058 &0.056 &0.055\\ \hline
5-9   &0.094 &0.082 &0.075 &0.079 &0.078 &0.078 &0.072 &0.069\\ \hline
10-14 &0.107 &0.094 &0.082 &0.075 &0.079 &0.079 &0.078 &0.073\\ \hline
15-19 &0.101 &0.104 &0.092 &0.080 &0.074 &0.078 &0.078 &0.077\\ \hline
20-24 &0.082 &0.091 &0.095 &0.083 &0.073 &0.068 &0.072 &0.073\\ \hline
25-29 &0.070 &0.081 &0.088 &0.090 &0.078 &0.070 &0.064 &0.068\\ \hline
30-34 &0.059 &0.068 &0.078 &0.085 &0.088 &0.077 &0.069 &0.063\\ \hline
35-39 &0.053 &0.056 &0.064 &0.075 &0.082 &0.085 &0.075 &0.067\\ \hline
40-44 &0.055 &0.052 &0.055 &0.063 &0.073 &0.079 &0.082 &0.072\\ \hline
45-49 &0.057 &0.053 &0.050 &0.053 &0.060 &0.069 &0.075 &0.078\\ \hline
50-54 &0.055 &0.054 &0.050 &0.047 &0.050 &0.057 &0.066 &0.072\\ \hline
55-59 &0.050 &0.050 &0.050 &0.047 &0.044 &0.047 &0.054 &0.062\\ \hline
60-64 &0.046 &0.045 &0.045 &0.046 &0.043 &0.040 &0.043 &0.050\\ \hline
65-69 &0.039 &0.039 &0.039 &0.039 &0.040 &0.037 &0.035 &0.038\\ \hline
70-74 &0.028 &0.030 &0.031 &0.031 &0.031 &0.032 &0.030 &0.030\\ \hline
75-79 &0.019 &0.020 &0.022 &0.023 &0.023 &0.023 &0.024 &0.023\\ \hline
80-84 &0.012 &0.012 &0.012 &0.013 &0.014 &0.014 &0.015 &0.016\\ \hline
85+   &0.007 &0.007 &0.008 &0.009 &0.009 &0.010 &0.010 &0.011\\ \hline \hline
\end{tabular}
\end{center}
\end{table}

\begin{table}
\begin{center}{Table S11. Prostate Cancer Mortality Rate ($10^{-5}$) during 1970 -- 2009 in New Jersey}
\begin{tabular}{rrrrrrrrr}
Age &1970-74&1975-79&1980-84&1985-89&1990-94&1995-99&2000-04&2005-09\\ \hline \hline
1-4   &0.00   &0.00   &0.00   &0.00   &0.00   &0.00   &0.00   &0.00\\ \hline
5-9   &0.00   &0.00   &0.00   &0.00   &0.00   &0.00   &0.00   &0.00\\ \hline
10-14   &0.00   &0.00   &0.00   &0.00   &0.00   &0.00   &0.00   &0.00\\ \hline
15-19   &0.00   &0.00   &0.00   &0.00   &0.00   &0.00   &0.00   &0.00\\ \hline
20-24   &0.00   &0.00   &0.00   &0.00   &0.00   &0.00   &0.00   &0.00\\ \hline
25-29   &0.00   &0.00   &0.00   &0.00   &0.00   &0.00   &0.00   &0.00\\ \hline
30-34   &0.00   &0.00   &0.00   &0.00   &0.00   &0.00   &0.00   &0.00\\ \hline
35-39   &0.00   &0.00   &0.00   &0.00   &0.00   &0.00   &0.00   &0.00\\ \hline
40-44   &0.00   &0.00   &0.00   &0.00   &0.00   &0.00   &0.00   &0.00\\ \hline
45-49   &1.24   &0.00   &1.72   &0.00   &1.07   &0.71   &1.01   &1.39\\ \hline
50-54   &4.12   &4.74   &4.74   &4.99   &3.38   &4.22   &2.99   &2.49\\ \hline
55-59  &11.51  &10.19  &12.28  &15.85  &14.74  &11.77   &8.64   &8.08\\ \hline
60-64  &34.03  &33.37  &35.28  &37.12  &39.01  &33.92  &23.68  &22.07\\ \hline
65-69  &66.53  &81.69  &75.08  &83.62  &86.97  &71.40  &54.16  &41.72\\ \hline
70-74 &143.24 &146.82 &145.70 &164.11 &166.76 &139.84 &101.66  &80.57\\ \hline
75-79 &263.56 &265.84 &271.33 &268.44 &292.58 &248.27 &181.05 &146.20\\ \hline
80-84 &363.80 &431.40 &427.61 &447.25 &502.60 &402.65 &334.34 &238.08\\ \hline
85+   &532.54 &550.08 &673.93 &684.08 &810.78 &756.29 &658.09 &535.58\\ \hline \hline
\end{tabular}
\end{center}
\end{table}

\begin{table}
\begin{center}{Table S12. Population Proportions during 1970 -- 2009 in New Jersey} \begin{tabular}{rrrrrrrrr}
Age &1970-74&1975-79&1980-84&1985-89&1990-94&1995-99&2000-04&2005-09\\ \hline \hline
1-4   &0.067 &0.056 &0.054 &0.057 &0.063 &0.060 &0.056 &0.054\\ \hline
5-9   &0.096 &0.082 &0.069 &0.068 &0.071 &0.077 &0.074 &0.069\\ \hline
10-14 &0.106 &0.097 &0.083 &0.069 &0.068 &0.071 &0.077 &0.074\\ \hline
15-19 &0.095 &0.100 &0.094 &0.080 &0.068 &0.067 &0.069 &0.074\\ \hline
20-24 &0.075 &0.083 &0.088 &0.086 &0.072 &0.062 &0.062 &0.064\\ \hline
25-29 &0.069 &0.078 &0.084 &0.089 &0.084 &0.071 &0.065 &0.065\\ \hline
30-34 &0.060 &0.071 &0.080 &0.088 &0.094 &0.087 &0.075 &0.066\\ \hline
35-39 &0.057 &0.060 &0.072 &0.081 &0.087 &0.092 &0.084 &0.074\\ \hline
40-44 &0.062 &0.056 &0.060 &0.070 &0.077 &0.082 &0.087 &0.081\\ \hline
45-49 &0.065 &0.059 &0.053 &0.057 &0.065 &0.072 &0.078 &0.083\\ \hline
50-54 &0.063 &0.062 &0.056 &0.050 &0.052 &0.060 &0.067 &0.073\\ \hline
55-59 &0.055 &0.057 &0.056 &0.051 &0.044 &0.046 &0.054 &0.062\\ \hline
60-64 &0.045 &0.047 &0.050 &0.049 &0.044 &0.039 &0.041 &0.048\\ \hline
65-69 &0.033 &0.036 &0.039 &0.041 &0.039 &0.036 &0.032 &0.035\\ \hline
70-74 &0.023 &0.025 &0.028 &0.030 &0.032 &0.031 &0.029 &0.027\\ \hline
75-79 &0.015 &0.016 &0.018 &0.020 &0.022 &0.023 &0.023 &0.022\\ \hline
80-84 &0.009 &0.009 &0.010 &0.011 &0.012 &0.014 &0.015 &0.016\\ \hline
85+   &0.005 &0.006 &0.006 &0.007 &0.007 &0.009 &0.010 &0.012\\ \hline \hline
\end{tabular}
\end{center}
\end{table}

\begin{table}
\begin{center}{Table S13. Prostate Cancer Mortality Rate ($10^{-5}$) during 1970 -- 2009 in New York}
\begin{tabular}{rrrrrrrrr}
Age &1970-74&1975-79&1980-84&1985-89&1990-94&1995-99&2000-04&2005-09\\ \hline \hline
1-4   &0.00   &0.00   &0.00   &0.00   &0.00   &0.00   &0.00   &0.00\\ \hline
5-9   &0.00   &0.00   &0.00   &0.00   &0.00   &0.00   &0.00   &0.00\\ \hline
10-14   &0.00   &0.00   &0.00   &0.00   &0.00   &0.00   &0.00   &0.00\\ \hline
15-19   &0.00   &0.00   &0.00   &0.00   &0.00   &0.00   &0.00   &0.00\\ \hline
20-24   &0.00   &0.00   &0.00   &0.00   &0.00   &0.00   &0.00   &0.00\\ \hline
25-29   &0.00   &0.00   &0.00   &0.00   &0.00   &0.00   &0.00   &0.00\\ \hline
30-34   &0.00   &0.00   &0.00   &0.00   &0.00   &0.00   &0.00   &0.00\\ \hline
35-39   &0.00   &0.00   &0.00   &0.00   &0.00   &0.00   &0.00   &0.00\\ \hline
40-44   &0.00   &0.00   &0.00   &0.00   &0.00   &0.00   &0.00   &0.00\\ \hline
45-49   &1.56   &0.85   &1.18   &1.54   &1.06   &1.10   &0.77   &0.81\\ \hline
50-54   &3.51   &4.07   &4.54   &4.95   &3.99   &4.43   &2.95   &2.95\\ \hline
55-59  &12.12  &11.66  &12.80  &14.12  &12.36  &12.18  &10.38   &8.57\\ \hline
60-64  &30.82  &31.84  &33.03  &36.15  &36.27  &29.65  &24.04  &19.73\\ \hline
65-69  &72.25  &72.44  &73.20  &81.53  &85.45  &68.30  &54.74  &44.04\\ \hline
70-74 &136.83 &137.86 &142.83 &154.51 &158.62 &137.89 &105.21  &83.94\\ \hline
75-79 &229.50 &245.46 &261.62 &255.64 &275.09 &235.81 &188.74 &150.54\\ \hline
80-84 &362.86 &394.83 &389.17 &417.67 &443.38 &370.03 &323.14 &253.43\\ \hline
85+   &425.94 &518.47 &602.90 &604.83 &750.84 &710.66 &637.20 &515.18\\ \hline \hline
\end{tabular}
\end{center}
\end{table}

\begin{table}
\begin{center}{Table S14. Population Proportions during 1970 -- 2009 in New York} \begin{tabular}{rrrrrrrrr}
Age &1970-74&1975-79&1980-84&1985-89&1990-94&1995-99&2000-04&2005-09\\ \hline \hline
1-4   &0.067 &0.057 &0.057 &0.059 &0.064 &0.061 &0.054 &0.052\\ \hline
5-9   &0.093 &0.081 &0.070 &0.071 &0.072 &0.078 &0.072 &0.065\\ \hline
10-14 &0.102 &0.094 &0.082 &0.070 &0.070 &0.073 &0.077 &0.071\\ \hline
15-19 &0.095 &0.099 &0.093 &0.082 &0.070 &0.072 &0.074 &0.077\\ \hline
20-24 &0.080 &0.087 &0.092 &0.089 &0.080 &0.069 &0.071 &0.074\\ \hline
25-29 &0.073 &0.082 &0.087 &0.092 &0.087 &0.077 &0.069 &0.072\\ \hline
30-34 &0.062 &0.072 &0.082 &0.088 &0.092 &0.085 &0.076 &0.067\\ \hline
35-39 &0.056 &0.059 &0.069 &0.080 &0.084 &0.087 &0.081 &0.072\\ \hline
40-44 &0.060 &0.055 &0.058 &0.067 &0.075 &0.079 &0.082 &0.077\\ \hline
45-49 &0.061 &0.056 &0.051 &0.054 &0.061 &0.070 &0.074 &0.078\\ \hline
50-54 &0.059 &0.059 &0.054 &0.048 &0.051 &0.058 &0.066 &0.071\\ \hline
55-59 &0.053 &0.054 &0.054 &0.049 &0.044 &0.045 &0.053 &0.062\\ \hline
60-64 &0.046 &0.046 &0.048 &0.046 &0.043 &0.039 &0.041 &0.048\\ \hline
65-69 &0.036 &0.037 &0.038 &0.039 &0.037 &0.035 &0.033 &0.035\\ \hline
70-74 &0.026 &0.027 &0.028 &0.028 &0.029 &0.030 &0.029 &0.027\\ \hline
75-79 &0.017 &0.018 &0.019 &0.020 &0.021 &0.022 &0.023 &0.022\\ \hline
80-84 &0.010 &0.011 &0.011 &0.011 &0.012 &0.013 &0.015 &0.016\\ \hline
85+   &0.005 &0.006 &0.007 &0.008 &0.008 &0.009 &0.010 &0.012\\ \hline \hline
\end{tabular}
\end{center}
\end{table}

\clearpage
\newpage

\begin{figure}[h]
\rule{2em}{0em}\centering{\psfig{file=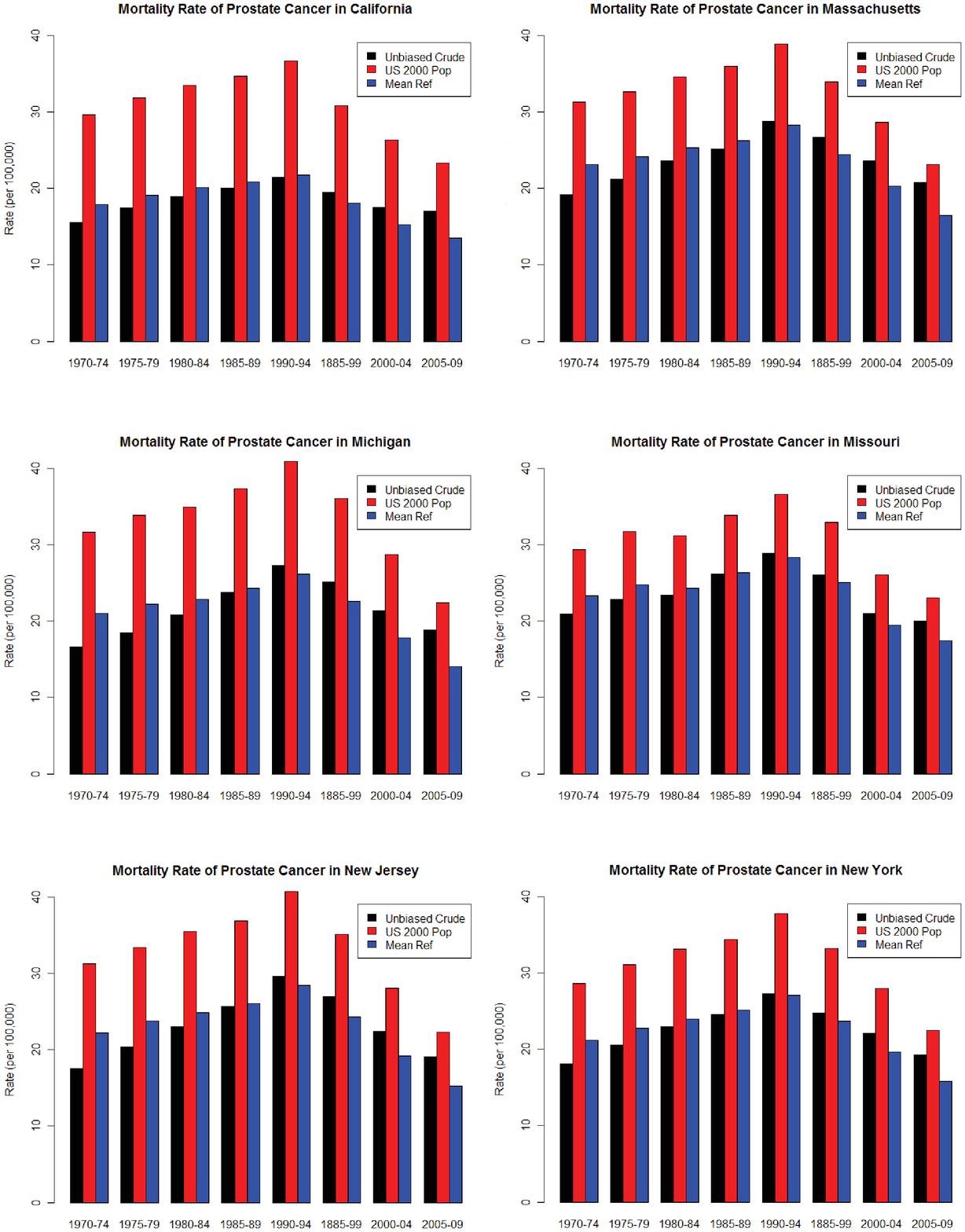,width=5in,height=7.5in}}
\begin{center}
{Figure S1. Comparison of age-adjusted mortality rates using the US Year 2000 Population Standard and the mean reference population to the crude rate of prostate cancer in six states during 1970-2009.}
\end{center}
\end{figure}

\begin{figure}[h]
\rule{2em}{0em}\centering{\psfig{file=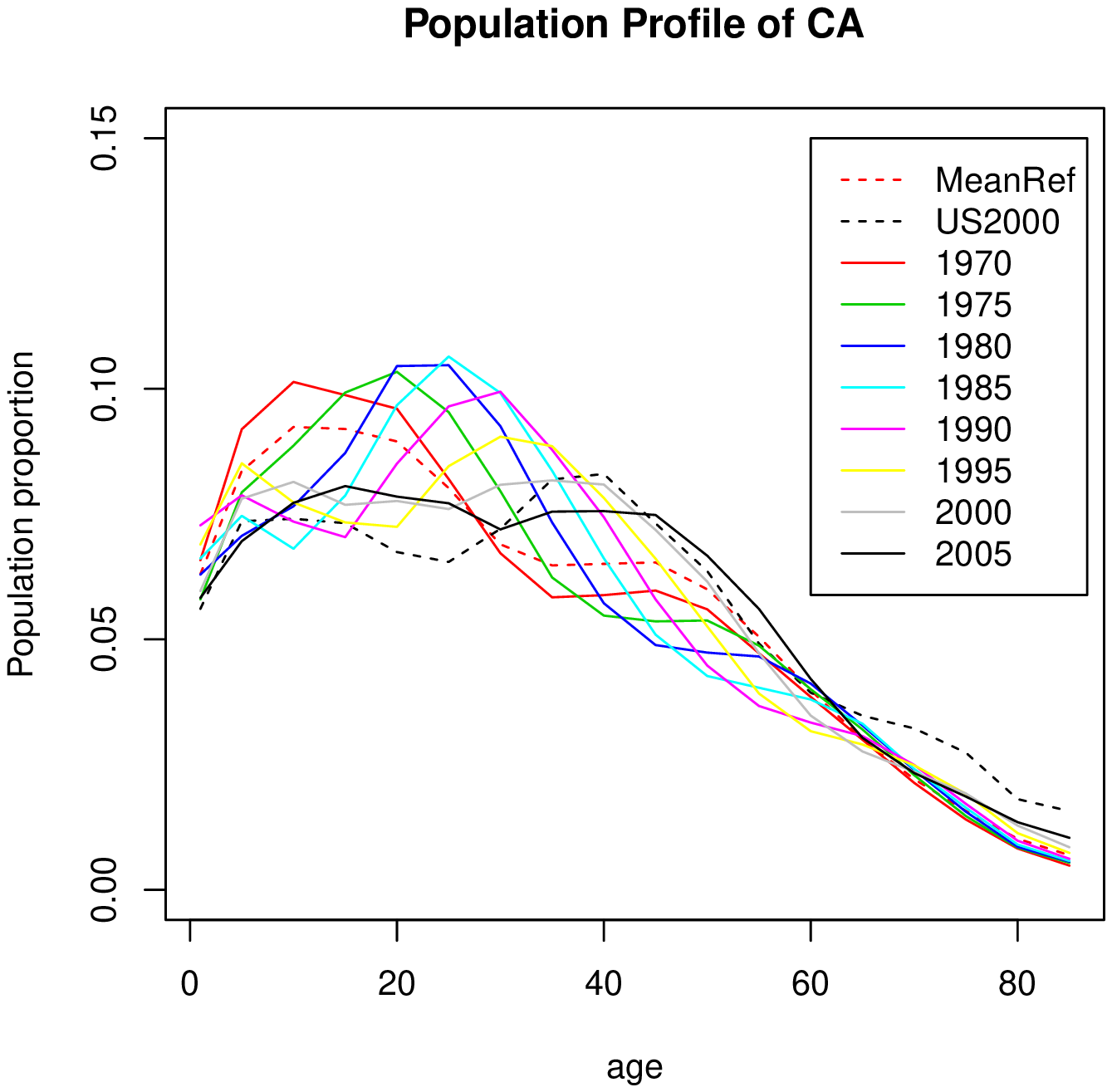,width=5.5in,height=5in}}
\begin{center}
{Figure S2. Age profile of population  in California during 1970--2009.}
\end{center}
\end{figure}

\begin{figure}[h]
\rule{2em}{0em}\centering{\psfig{file=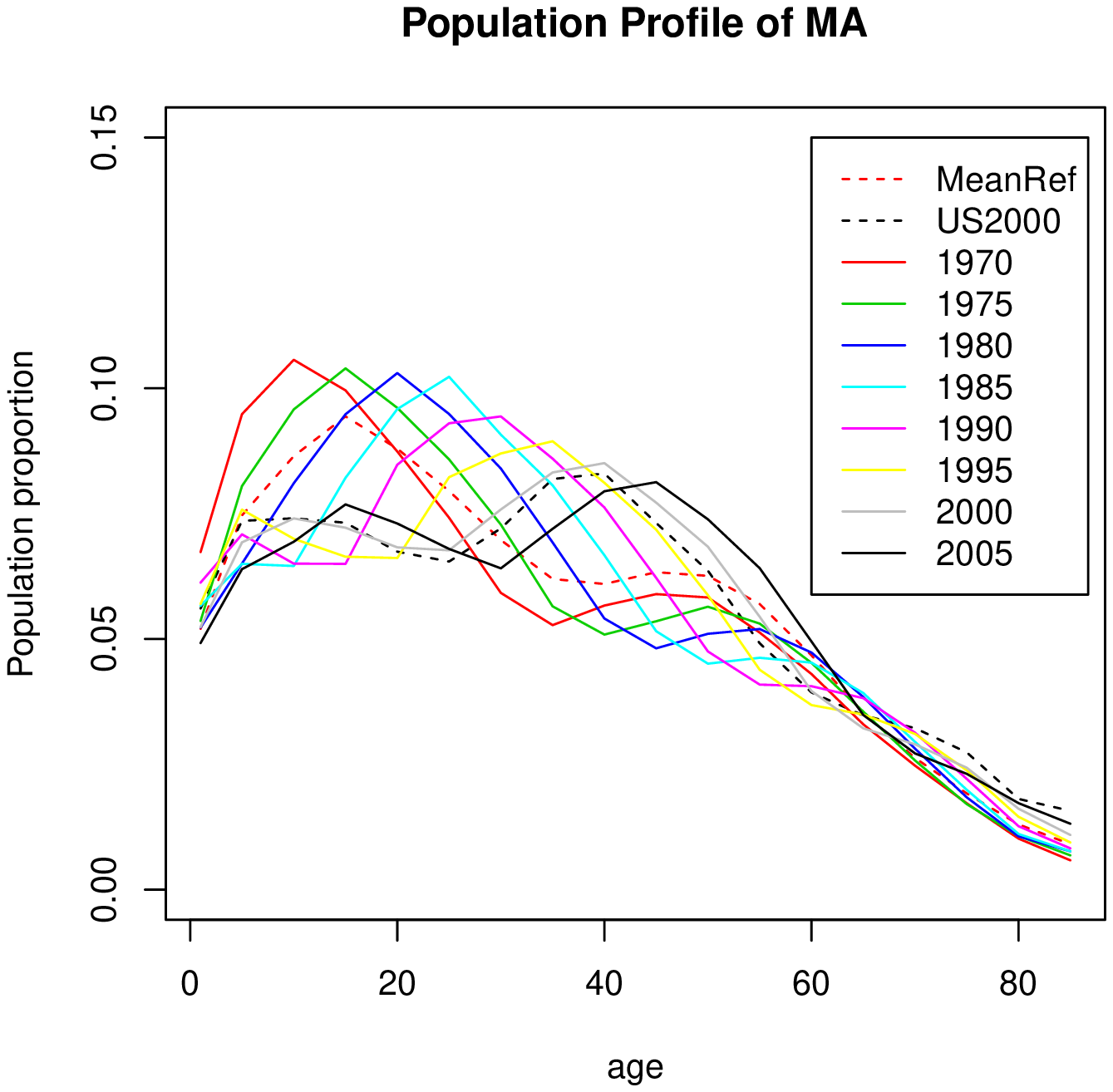,width=5.5in,height=5in}}
\begin{center}
{Figure S3. Age profile of population  in Massachusetts during 1970--2009.}
\end{center}
\end{figure}

\begin{figure}[h]
\rule{2em}{0em}\centering{\psfig{file=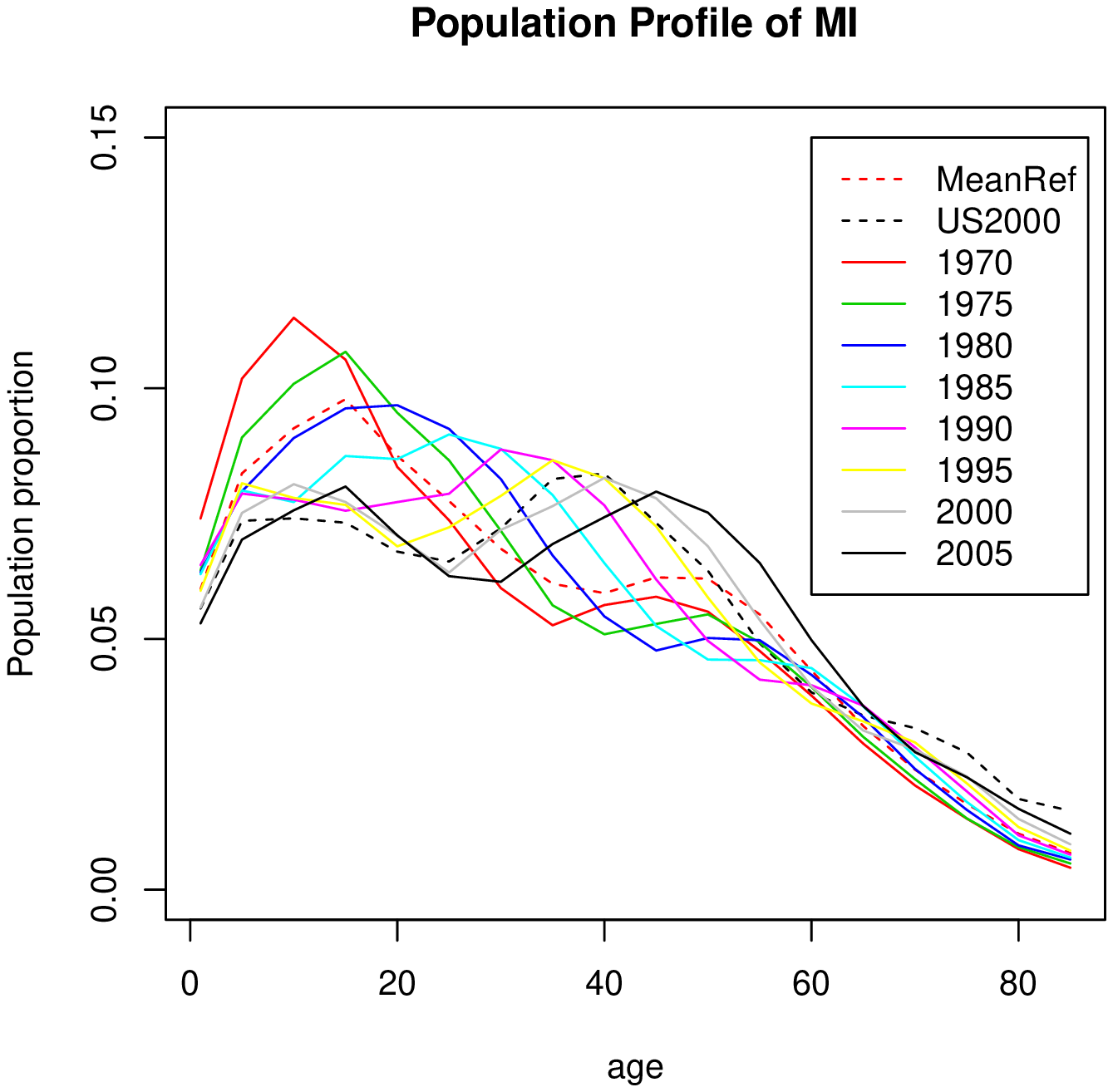,width=5.5in,height=5in}}
\begin{center}
{Figure S4. Age profile of population  in Michigan during 1970--2009.}
\end{center}
\end{figure}

\begin{figure}[h]
\rule{2em}{0em}\centering{\psfig{file=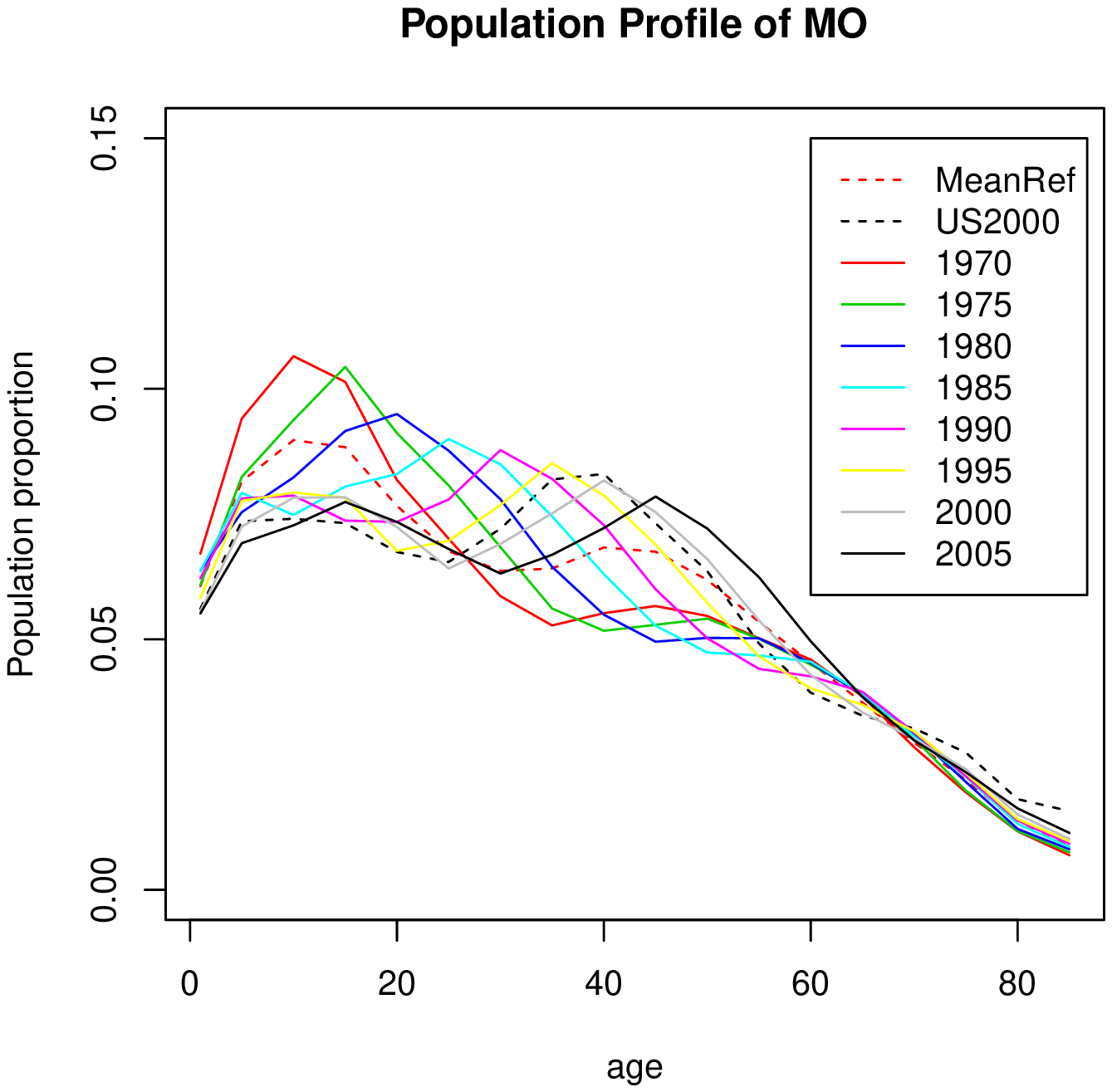,width=5.5in,height=5in}}
\begin{center}
{Figure S5. Age profile of population  in Missouri during 1970--2009.}
\end{center}
\end{figure}

\newpage

\begin{figure}[h]
\rule{2em}{0em}\centering{\psfig{file=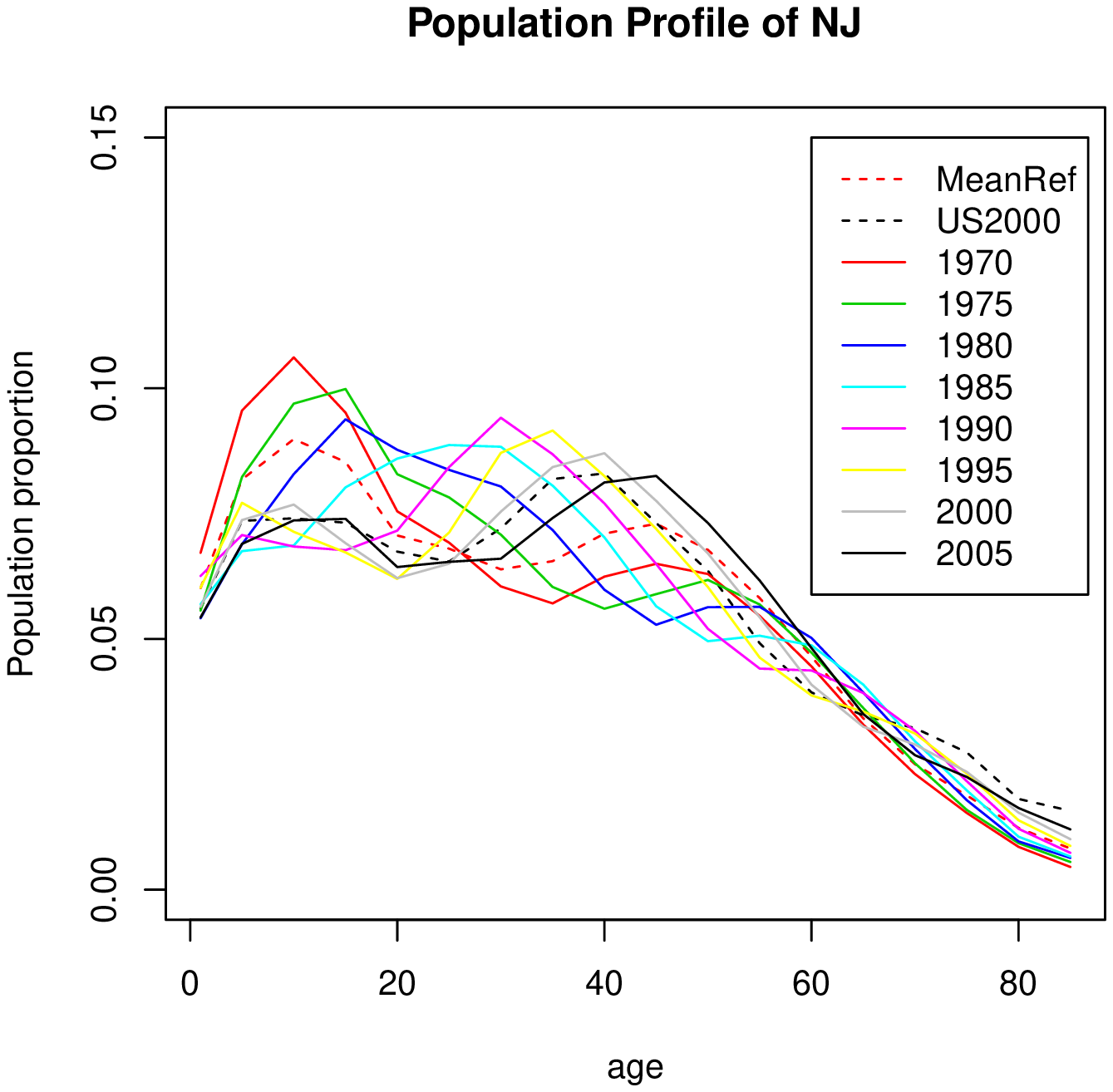,width=5.5in,height=5in}}
\begin{center}
{Figure S6. Age profile of population  in New Jersey during 1970--2009.}
\end{center}
\end{figure}

\begin{figure}[h]
\rule{2em}{0em}\centering{\psfig{file=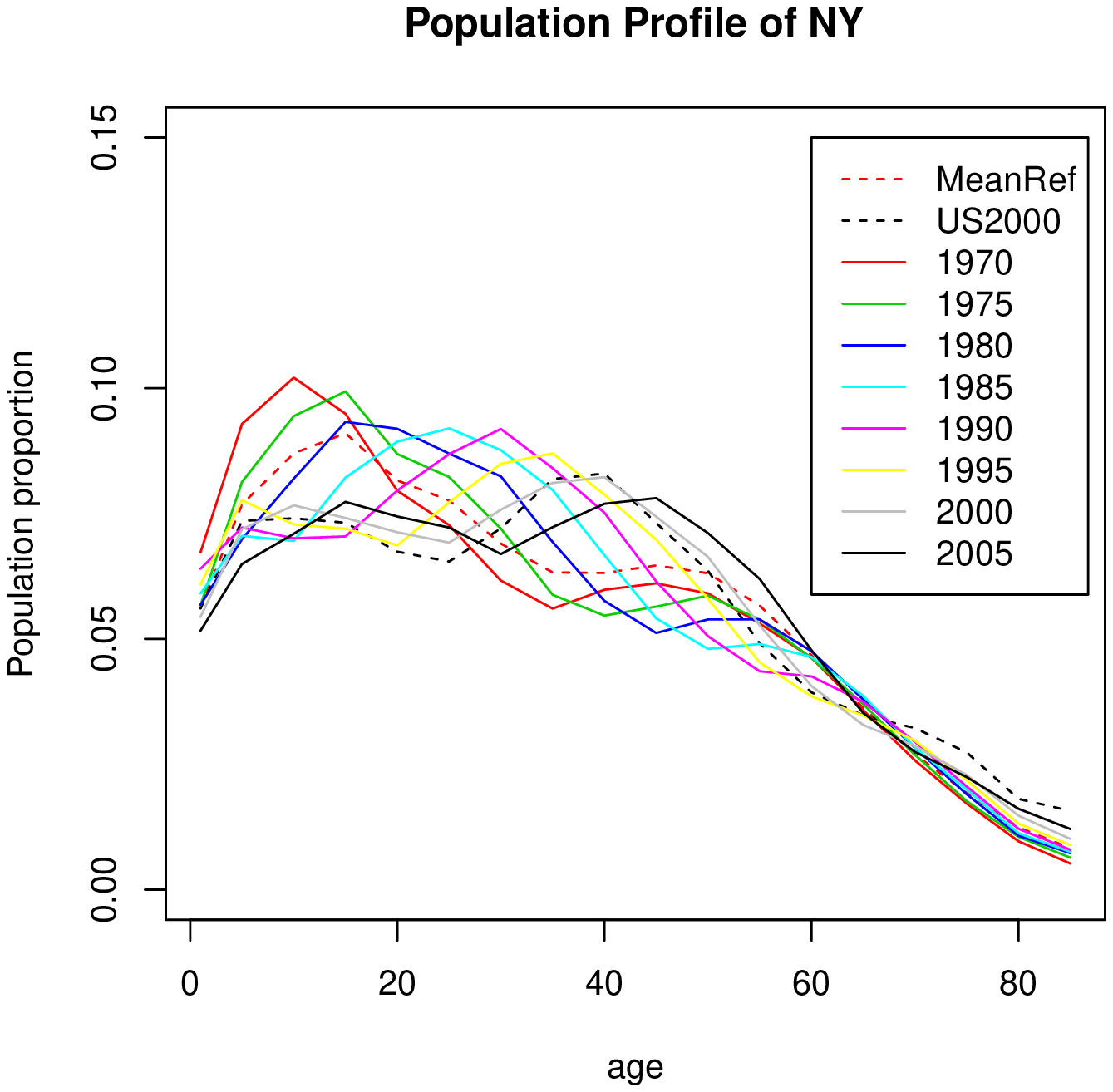,width=5.5in,height=5in}}
\begin{center}
{Figure S7. Age profile of population  in New York during 1970--2009.}
\end{center}
\end{figure}

\end{document}